\documentclass[12pt]{article}

\usepackage{amsfonts, latexsym, amsbsy, amssymb, amsmath}
\usepackage{leftidx}

\usepackage{graphicx}

\usepackage[T1]{fontenc}
\usepackage[ngerman,english]{babel}
\usepackage[utf8]{inputenc}

\usepackage[authoryear]{natbib}
\title{\bf{The Spectrum of He$^+$ as a Proving Ground for Bohr's Model of the Atom: A Legacy of Williamina Fleming's Astrophysical Discovery}}
\author{Maria McEachern$^a$ and Bretislav Friedrich$^b$}
\date{%
    $^a$\emph{Center for Astrophysics $\vert$ Harvard \& Smithsonian, 60 Garden St., Cambridge, MA 02138, U.S.A.}\\%
    $^b$\emph{Fritz-Haber-Institut der Max-Planck-Gesellschaft, \\ Faradayweg 4-6, D-14195 Berlin, Germany}
  \\[2ex]%
    \today}
    
\begin{document}
\maketitle
	\newpage
	\begin{abstract}
In 1896, Edward Charles Pickering (1846-1919), Director of the Harvard College Observatory (HCO), reported in a trio of publications the observation of ``peculiar spectra'' of the southern star $\zeta$ Puppis, which he  attributed to an ``element not yet found in other stars or on earth.'' Supported by laboratory spectra obtained by Alfred Fowler (1868-1940), Niels Bohr (1885-1962) showed in 1913 that this ``element'' was in fact ionized helium, He$^+$. Its spectrum has become known as the Pickering Series, even though Pickering credited Williamina Fleming (1857-1911) for the discovery. Fleming was one of HCO's ``computers'' and the future Curator of the Astronomical Photographic Glass Plate Collection. The series of spectral lines associated with Pickering's name played a unique role on the path to quantum mechanics by serving as a proving ground for Bohr's model of the atom. Our examination of the discovery of the Pickering series relied on the records held at the Center for Astrophysics $\vert$ Harvard \& Smithsonian (the successor institution to HCO), especially the Notebooks and Diaries of Williamina Fleming and others as well as on the Center's Glass Plate Collection. Glimpses of the ``peculiar sociology'' of a research institution, half of whose staff were women employed on grossly unequal terms with men, are given in the course of the narrative. 
	\end{abstract}
	
	\section{Prologue}
	\label{Prelude}
	
	The Bohr model of the atom was conceived at a time characterized by Abraham Pais\footnote{Pais borrowed these words from \emph{A Tale of Two Cities} by Charles Dickens.} as ``the epoch of belief, ... the epoch of incredulity''      \citep{Pais_1986}, p. 211. The model entailed the ``fourth coming'' of Planck's constant, following upon its previous appearances in Planck's black body radiation law, Einstein's light quantum hypothesis, and Einstein's quantum treatment of the heat capacity of solids. See, e.g., \citep{Kragh_2002}. 
	
	Although the Bohr model had its precursors in the work of Arthur Erich Haas \citep{Haas_1910,Wiescher_2021} and John William Nicholson \citep{Nicholson_1911}, it was Bohr's 1913 postulates concerning the electron dynamics in a planetary atom that led to the explanation of the Balmer series in terms of transitions between stationary electronic states and an interpretation of the Rydberg constant of atomic hydrogen in terms of fundamental constants \citep{Bohr_1913He}. However, the model remained tentative until the advent of quantum mechanics in 1925-1926 and was regarded as \emph{ad hoc} by Bohr's contemporaries as it was tailored for atomic hydrogen. Then, Bohr thought of the Pickering Series, as the spectrum of the He$^+$ atomic ion was known, and recognized that his model should be as applicable to He$^+$ as it was to atomic hydrogen. But was it? 
	
\begin{figure}
		\centering
		\includegraphics[scale=0.5]{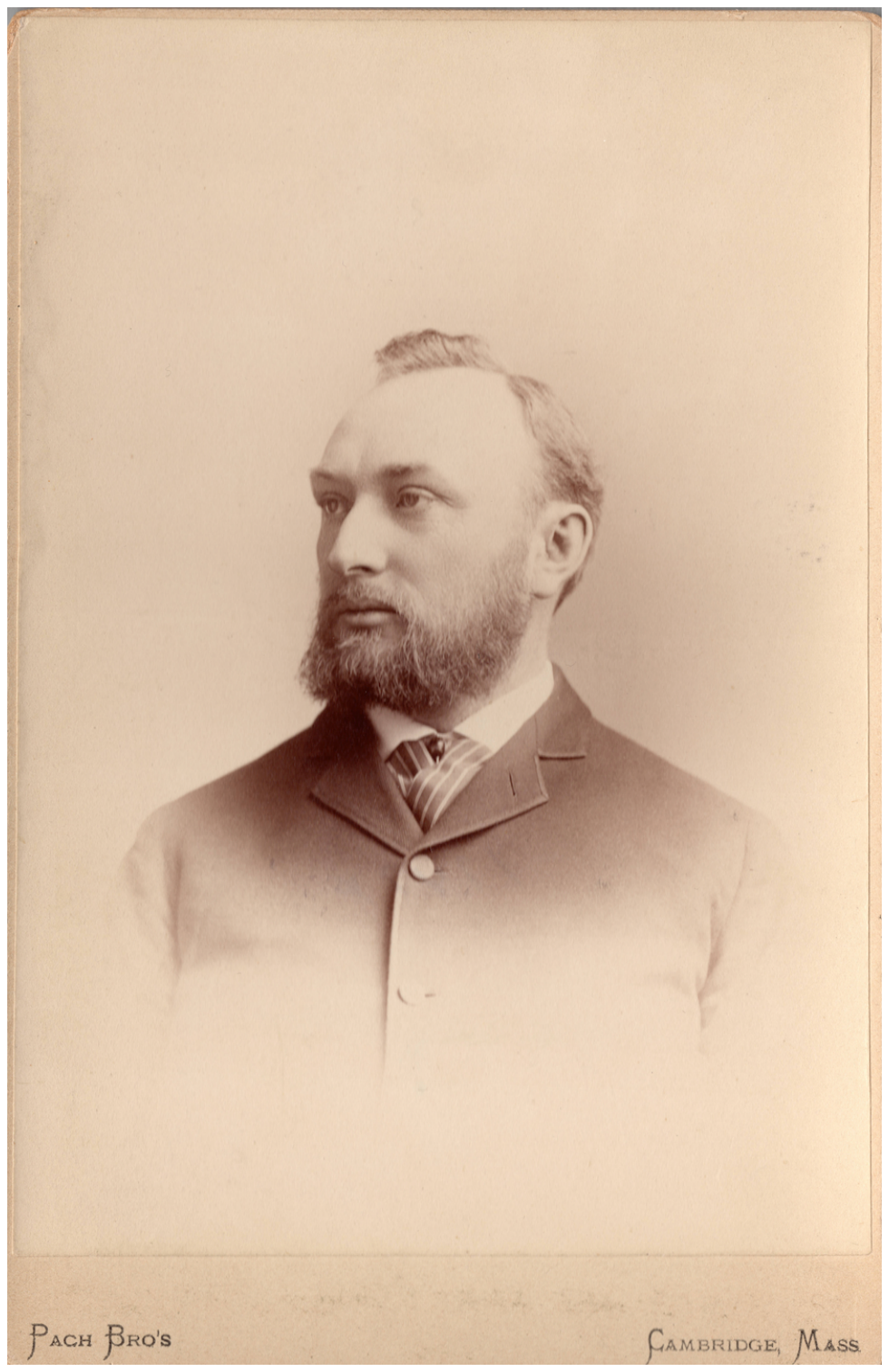} 
		\caption{\label{fig:Pickering_1891-1896} Edward Charles Pickering (1846-1919). Photo taken between 1891 and 1896 at the Pach Brothers studio in Cambridge, Massachusetts. Courtesy of the Harvard College Observatory.} 
	\end{figure}

\begin{figure}
\centering
\includegraphics[scale=0.6]{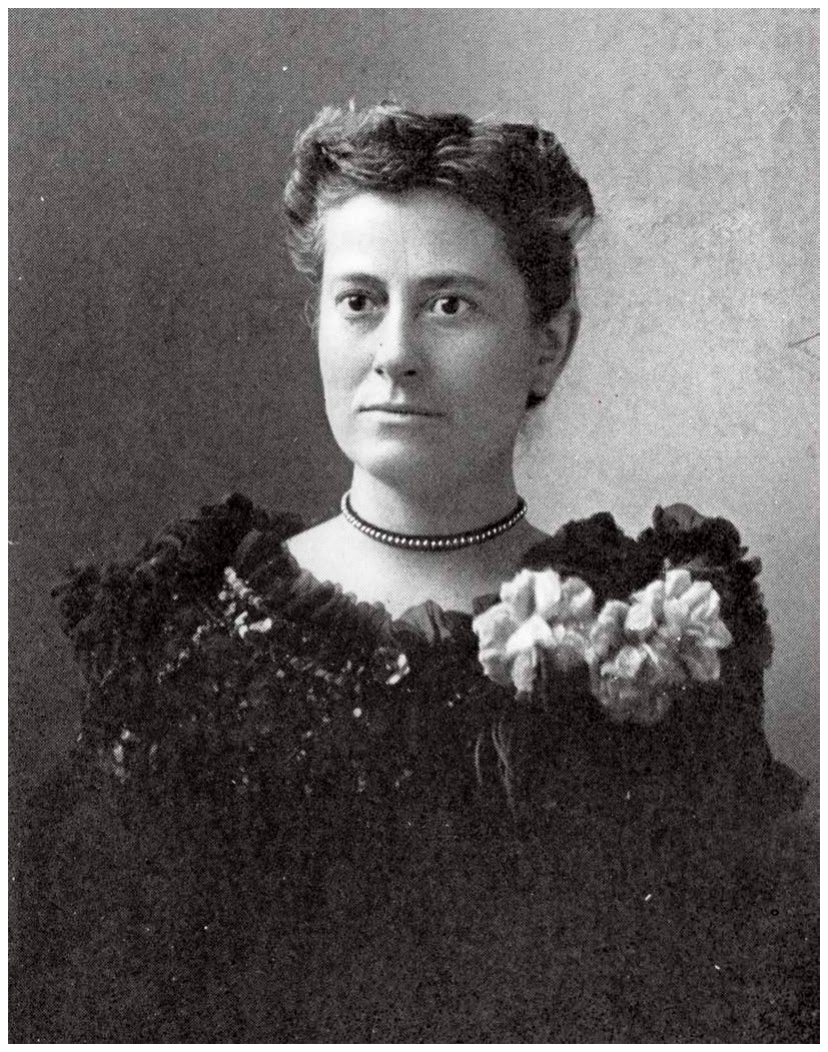} 
\caption{\label{fig:Williamina_Fleming_7006-31} Williamina Paton Stevens Fleming (1857-1911). Courtesy of the Harvard College Observatory.}
\end{figure}

	In this Chapter, we describe the history of the discovery of the spectrum of He$^+$ and show how it served as a proving ground for Bohr's model of the one-electron atom. In the process, we demonstrate that the ``Pickering Series'' had in fact been discovered by one of the women who worked as a ``computer'' at the Astronomical Observatory of Harvard College (or Harvard College Observatory for short and abbreviated as HCO) that was headed at the time by Edward Charles Pickering, Fig. \ref{fig:Pickering_1891-1896}.  This astronomer and stellar spectroscopist who worked under him was Williamina Paton Fleming, Fig. \ref{fig:Williamina_Fleming_7006-31}. 
	
	Although the postulates on which Bohr built his atomic model remained incomprehensible until the discovery of quantum mechanics, the model's credibility was boosted by its ability to explain -- with a five-digit accuracy -- the Pickering Series.
		
	This Chapter is structured as follows: In Section \ref{Protagonists}, we provide brief biographies of the protagonists of this Chapter, Edward Pickering and Williamina Fleming. 
Section \ref{Stellar} introduces the workings of the Harvard College Observatory during the Pickering era (1877-1919). Subsections \ref{Photometry} and \ref{Mapping} detail the project of mapping the starry skies by means of astrophotography and its spectrally-resolved variant.  Subsections \ref{Classification} and \ref{Variable} introduce stellar classifications and variable stars. In Section \ref{Zeta}, we explore the holdings of the Harvard College Observatory, including its collection of photographic glass plates, and find evidence that the Pickering Series was actually discovered by Williamina Fleming as a ``peculiar spectrum'' in the spectrally-resolved photographs of the southern variable star $\zeta$ Puppis. Section \ref{HePlus} describes how the spectrum of He$^+$ became a proving ground for Bohr's model of the one-electron atom and what impression the model's extended applicability to a species for which it had not been tailored left on the community. Finally, in Section \ref{Conclusions}, we comment on the significance of He$^+$ for astronomy itself as well as for astrochemistry. 	
	
	\section{The Protagonists}
	\label{Protagonists}
	
	\subsection{Edward Charles Pickering}
	\label{Pickering}

Edward Charles Pickering was born in Boston on 19 July 1846 into a patrician family. Upon graduating from the Lawrence Scientific School of Harvard University\footnote{A forerunner of what is today Harvard's School of Engineering and Applied Sciences.} at age nineteen and a stint as an Instructor of Mathematics at his \emph{alma mater}, he was hired in 1867 by the Massachusetts Institute of Technology as a Professor of Physics. During the following decade, Pickering developed MIT's Physical Laboratory, the first such establishment in the U.S. \citep{Pickering_1869}. The way the laboratory was set up and administered reflected Pickering's credo: ``There are no secrets in Science'' \citep{AJC_1919}. The laboratory was open not only to students and teachers of physics but also to the lay public wishing to carry out their research. In 1876, Pickering was appointed by Harvard President Charles Eliot (1834-1926) as the fourth Director of the Harvard College Observatory (HCO). 

Pickering entered his duties at the HCO in 1877 with a plan to investigate the brightness of stars by photometry. In 1882, he extended traditional photometric techniques by launching a program in \emph{photographic astronomy}. This innovation was based on dry photographic plates (i.e., glass plates coated with a gelatin emulsion of silver bromide that could be stored before exposure and developed at leisure after exposure). In 1886, he launched the Henry Draper Memorial project of spectrally-resolved photographic mapping of both the northern and southern skies. The telescopes used entailed a dispersive element (an ocular prism) in order to spectrally resolve the star images recorded as structured smudges on the photographic glass plates. Pickering's predilection for spectroscopy originated in his training as a physicist. The spectrally-resolved astrophotography enabled an unprecedented accumulation of empirical data intended to prepare the soil for a better understanding of the nature of stars. Moreover, given that the astrophotographs were taken at known times (the expositions lasted minutes to tens of minutes), a set of subsequent astrophotographs contained information about the variation in time of the stars' brightness and spectra. As a result, the study of \emph{variable stars} had become a major preoccupation of the astrophysical community. Enhanced sensitivity to time variation was achieved by the technique of superposing a negative and a positive taken at different times, leading to the discovery of almost 3,500 variables at the HCO during Pickering's tenure.

Pickering cultivated relations with potential sponsors of the HCO such as Mary Anna Draper (1839-1914), Catherine Bruce (1816-1900), and Uriah Boyden (1804-1879). Mrs. Draper would fund the Henry Draper Memorial catalog of stellar spectra, so named after her late husband, the physician and amateur astronomer Henry Draper (1837-1882), who pioneered astro-photography and, assisted by his wife, recorded stellar spectra as early as 1872. The Boyden fund made it possible to build HCO's permanent station in Arequipa, Peru, to map out the southern sky. It was at Arequipa where the spectrum of $\zeta$-Puppis was recorded. Miss Bruce's gift covered the cost of a double-telescope (Bruce Telescope) consisting of a tracking telescope that would be locked to a guiding star (to compensate for the star's apparent motion due to the earth's rotation) and a telescope through which the astro-photograph of a given star was taken. It was the Bruce Telescope that was installed at Arequipa. If need be, Pickering would also make financial contributions to the HCO from his own pocket \citep{Bailey_1932}.

In 1891, William Henry Pickering, the brother of Director Pickering, who had been hired as an assistant, was put in charge of the southern outpost in Arequipa. Much to Director Pickering's dismay, William was submitting his observations of the planet Mars via telegraph to the New York Herald in anticipation of the 1892 opposition of the Red Planet. These missives helped to stir the frenzy of the general public eager to receive any news regarding the Martian channels interpreted as canals created, perhaps, by the Martian public works ... There was great general interest in the potential for communication with the inhabitants of Mars as the upcoming opposition approached. It was, as journalistically described, ``the time of the great Mars boom, when public imbecility and journalistic enterprise combined to flood the papers and society with `news from Mars,' and queries concerning Mars, most exasperating to grave thinkers and workers in science'' \citep{Clerke_1896} and \citep{Shindell_2023}. Edward Pickering was furious and, given William's great liberties taken in ignoring the work which he had actually been assigned to do and his continually mounting extravagances, he recalled his brother with the approval of the Harvard Corporation. 

As noted by Annie Jump Cannon (1863-1941) \citep{AJC_1919}:
\begin{quote}
In the early days of photographic astronomy, Professor Pickering foresaw a great opportunity for woman's work, and gradually, from 1884, the staff of woman assistants was instituted [at the HCO], first for simple computing and examination of the photographs, then for posts of greater responsibility, until independent investigations were made by them.
\end{quote}
Williamina Fleming, Annie Jump Cannon, and Henrietta Swan Leavitt (1868-1921) were among the most prominent beneficiaries of Pickering's foresight. The exemplary partnership of Mary Anna and Henry Draper was likely suggestive of the possibilities of involving women in research.

As a man of the new Industrial Age, Pickering believed in the specialization of labor. The Women Astronomical Computers were chosen ``to work, not to think'' \citep{CPG_1979}, p. 54. Their work was to analyze and not to interpret the data derived from the astrophotographic glass plates. Thus, thanks to the growth of astronomical spectroscopy which enabled astronomical work in daylight \citep{Lankford_1997}, p. 310, the Harvard College Observatory came to embody the ``factory observatory'' \citep{Lankford_1997}, p. 309. In an address to the Harvard branch of Phi Beta Kappa, Pickering observed that \citep{Johnson_2005}, p. 18:
     \begin{quote}
A great observatory should be as carefully organized and administered as a railroad. Every expenditure should be watched, every real improvement introduced, advice from experts welcomed, and if good, followed, and every care taken to secure the greatest possible output for every dollar expenditure. A great savings may be effectuated by employing unskilled and therefore inexpensive labor, of course under careful supervision.
     \end{quote}
 Inexpensive labor indeed: the women's pay was typically 25 cents per hour \citep{Sobel_2016}, p. 113.

Pickering was the recipient of the highest honors from both the U.S. and European institutions; these included membership in the American Academy of Arts and Sciences (1867), the National Academy of Sciences (1873), as well as national societies of England, Germany, Ireland, Italy, Mexico, Russia, and Sweden. Moreover, he served as President of the American Astronomical Society (1905-1919) as well as of the American Association of Variable Star Observers (A.A.V.S.O.), founded in 1911, which coordinated observations and their analysis by mainly amateur astronomers \citep{AJC_1919}. He was also the founder and first President of the Appalachian Mountain Club. Edward Pickering died in Cambridge, Massachusetts on 3 February 1919. The International Astronomical Union named in 1935 a lunar crater \citep{PickeringLunarCrater} and in 1973 a Martian crater \citep{PickeringMartianCrater} after Pickering. In addition, Asteroid 784, discovered in 1914, was named Pickeringia \citep{Schmadel_2007}.

	\subsection{Williamina Fleming}
	\label{Fleming}

Williamina Paton Stevens Fleming was born in Dundee, Scotland, on May 15, 1857. She was one of several surviving children in her family at a time when child mortality accounted for nearly half of all deaths in Scotland \citep{ScotlandMortality}. Her father, a craftsman, was an early enthusiast of the recently discovered daguerreotype imaging. Williamina thus became familiar already as a child with this early photographic technique and its utility for recording images.
		
		Williamina became a teaching apprentice (pupil-teacher) at the age of fourteen and would continue on in this occupation for five more years. In 1877, she married James Orr Fleming and, in 1879, the couple followed other members of Williamina's family in leaving Scotland with the hope of starting a better life in the United States. A year later, Williamina found herself pregnant -- and abandoned by her husband. Left to fend for herself and her unborn child, she found employment in the household of the Director of the Harvard College Observatory, Edward Charles Pickering, where she was hired to provide household help along with copying assistance. Director Pickering and his wife, Lizzie Sparks Pickering, may have been impressed with her intelligence and facility with numbers in managing household accounts. Eventually, and quite possibly at the urging of Mrs. Pickering, Williamina began to work part-time at the Observatory as a copyist and computer. 
		
		Williamina, or, Mina, as she was familiarly known to friends and colleagues, returned home to Scotland to be with her family for the birth of her son, whom she named Edward Charles Pickering Fleming in gratitude for Pickering's kindness. In 1881, on her return to Cambridge, Massachusetts, Fleming began full-time employment at the HCO. She became a member of the regular staff, one of several who comprised the corps of women astronomical computers. Included among her duties was to oversee and edit several periodicals and other publications issued by the Observatory.  
		
		During her early career, she worked on the \emph{Harvard Photometry} project in which all of the stars visible from the Observatory were photographed and assigned a magnitude (a degree of brightness), relative to Polaris, the North Star, which stood as the standard. 
		
		Once the Henry Draper Catalog project was launched in 1886, Williamina Fleming became a full-fledged participant. She was involved in the investigation and cataloging of stellar spectra as recorded on photographic glass plates that held images of the night sky in both the Northern and Southern Hemispheres. Under her charge labored the corps of ``Women Astronomical Computers'' who, besides herself, inspected and analyzed the recorded images. Fleming organized the workflow and oversaw every phase of producing data gleaned from the plates. The data were published in multiple volumes of the \emph {Annals of the Astronomical Observatory of Harvard College}. Observations of objects of special note might be hurried to press in one of the Observatory’s \emph{Circulars}. Each issue of these titles was edited, proofread, prepared for publication and, following printing, inspected by Fleming. During the first three directorships of the Observatory, eight volumes of the \emph{Annals} had been published while, under the Pickering administration, this number rose to nearly one hundred. 		 		 
		
		In 1899, additional responsibility, and a corresponding job title, were bestowed upon Fleming. On Monday morning, January 16, 1899, \emph{The Boston Herald} carried the front-page headline ``Harvard Honors Women'' while a bold-print subheadline conveyed the news, ``Mrs. Fleming Appointed Curator of Astronomical Photographs, Charged With Their Care'' and, in less bold print, ``She is the First of Her Sex To Have Her Name Placed with the List of the Officers'' \citep{Herald_1899_1}, p. 1. The article itself went on to say \citep{Herald_1899_1}, p. 2: 
\begin{quote}
Mrs. Mina Fleming, the recently appointed Curator of Astronomical Photographs, has a worldwide reputation at once as a painstaking and patient investigator and as a brilliant discoverer in the field that is covered by the Henry Draper Memorial. The names of Mrs. Draper and Mrs. Fleming will go down in astronomical history in honorable conjunction with those of Caroline Herschel, Mary Somerville, and Maria Mitchell.  
\end{quote}   
      
      It is entirely appropriate that Williamina should have been named as the first Curator of Astronomical Photographs at the Harvard College Observatory. Besides her numerous discoveries made via the astrophotographic glass plates, thanks to her familiarity with daguerreotype images since childhood, Fleming knew and thoroughly trusted the photographic process. So much so that, in 1893,  she defended this new method of astronomical investigation against its detractors writing that \citep{Fleming_1893c}:
 \begin{quote}
 One must not always cling to the earliest method of accomplishing anything and assume that because it was the earliest and has held sway for centuries, it must consequently be the best, and also the only way.
 \end{quote} 
Annie Jump Cannon would later write in regard to Williamina's staunch defense of the use of photography in astronomical observation \citep{Cannon_1911}: 
 \begin{quote}
In the early days, when celestial photographs were rare, and some of these discoveries were attributed by skeptics to defects on the film, she never doubted the validity of the photographic evidence. Her industry was combined with great courage and independence.
 \end{quote}
    
      During that year, the Columbian Exposition was held in Chicago and the HCO was expected to participate and provide exhibition material, which would be prepared by Fleming.  Besides describing, in depth, the work at the Observatory, the many discoveries made by her colleagues (to whom she is not the least bit stingy in giving credit), and herself, she discussed, at some length, the great benefit of the use of photography in astronomy and its utility in studying the true elemental makeup of stars from their spectra.  Fleming wrote   \citep{Fleming_1893b} that astronomical photographs must be considered
\begin{quote}
more reliable for in the case of a visual observer, you have simply his statement of how the object appeared at a given time as seen by him alone, while here you have a photograph in which every star speaks for itself, and which can at any time, now or in the years to come, be compared with any other photographs of the same part of the sky.
\end{quote}
Fleming would also write a paper on women’s participation in astronomical study entitled ``A Field for Woman’s Work in Astronomy'' to be read at the Exposition and,  subsequently, published in the journal \emph{Astronomy and Astrophysics} \citep{Fleming_1893a}.

Given her numerous responsibilities, and, what must have been, at least at times, repetitive work, it is little wonder that, in her Diary of March 1900, which was kept as part of Harvard University's celebration of the New Millennium and would be placed into a time capsule, Williamina noted on the 8 March 1900 at 8:00 p.m., ``Miss Christopherson came to massage [my] hand and arm'' \citep{Fleming_1900}, p. 18.  She mused that, ``If one could only go on and on with original work, looking for new stars, variables, classifying spectra and studying their peculiarities and changes, life would be a most beautiful dream'' \citep{Fleming_1900}, p. 9; however, she added, ``but you come down to its realities when you have to put all that is interesting aside in order to use most of your available time preparing the work of others for publication'' \citep{Fleming_1900}, pp. 9-10. Still, she appreciated her role at the Observatory, ``contented to have such excellent opportunities for work … and proud to be considered of any assistance to such a thoroughly capable Scientific man as our Director [Pickering]'' \citep{Fleming_1900}, p. 10.

  \begin{figure}
		\centering
		\includegraphics[scale=0.83]{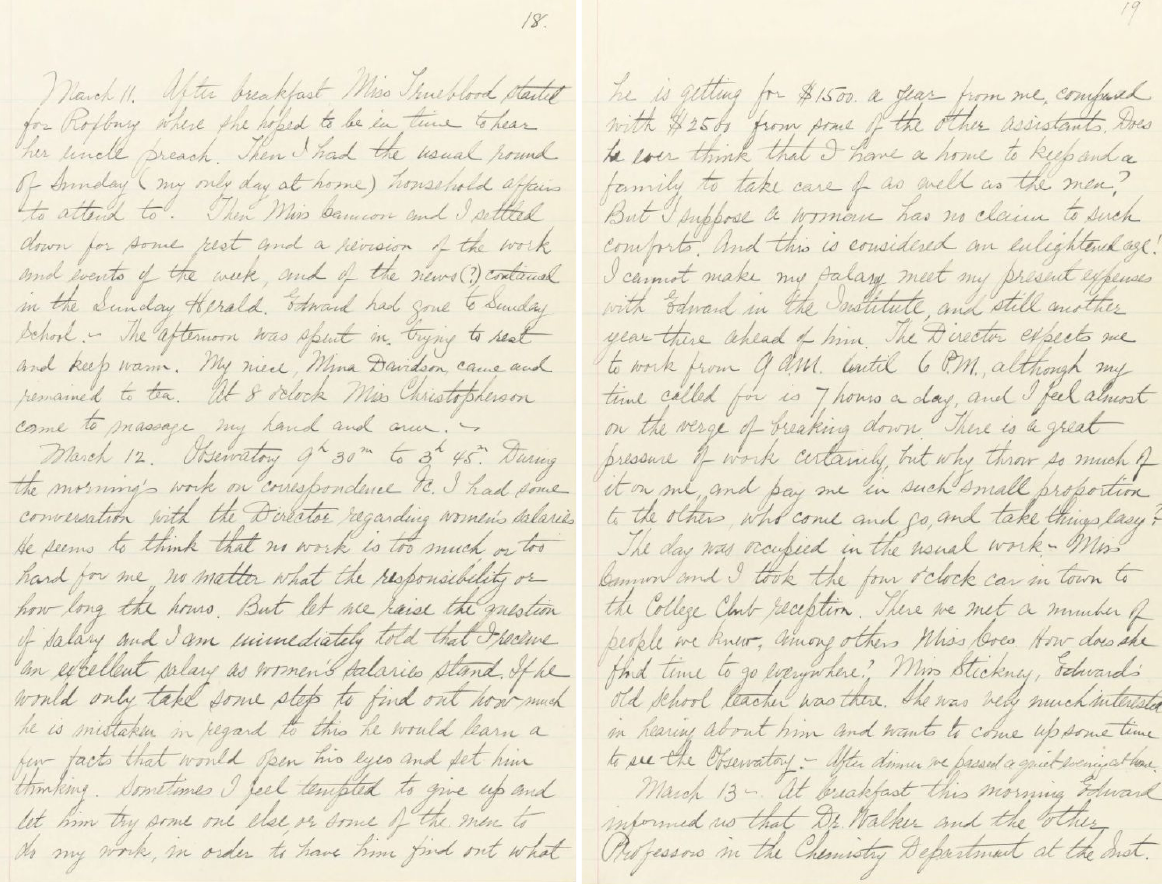} 
		\caption{\label{fig:Fleming_Journal_18-19} Pages 18 and 19 from Williamina Fleming's 1900 Journal \citep{Fleming_1900}. Williamina's annual salary of \$1,500 as Curator of Astronomical Photographs would correspond to about 50 cents per hour. For male assistants, the annual salary was \$2,500 according to this record. Courtesy of Harvard University Archives.} 
	\end{figure}

 On the subject of pay, Fleming felt the unfairness of her situation. In her 1900 journal \citep{Fleming_1900}, pp. 18-19, see also Fig. \ref{fig:Fleming_Journal_18-19}, she noted on March 12:
 \begin{quote}
I had some conversations with the Director regarding women's salaries. He seems to think that no work is too much or too hard for me, no matter what the responsibility or how long the hours are. But let me raise the question of salary and I am immediately told that I receive an excellent salary as women's salaries stand. ... Does he ever think that I have a home to keep and a family to take care of as well as the men? But I suppose a woman has no claim to such comforts. And this is considered an enlightened age! I cannot make my salary meet my present expenses with Edward [Williamina's son] in the Institute [MIT] and still another year there ahead of time.\footnote{Williamina's son, Edward Charles Pickering Fleming, was born in Dundee, Scotland, in 1879. After Williamina's departure for the U.S. in 1881, he was being cared for in Scotland by Williamina's mother. The boy joined Williamina in Cambridge, Massachusetts, when he was eight, accompanied by a female caretaker. He graduated from MIT in 1901 (in mining engineering) and was naturalized as a U.S. citizen in 1904. He died in California in 1962 \citep{Hughes_2012}, pp. 691-692.} The Director expects me to work from 9 a.m. to 6 p.m., although my time called for is 7 hours a day, and I feel almost on the verge of breaking down. There is a great pressure of work certainly, but why throw so much of it on me, and pay me in such small proportion to the others, who come and go, and take things easy?
 \end{quote}
   
   During her many years of active research, Fleming classified 10,351 stellar spectra and discovered more than 300 variable stars and 10 novae by recognizing the particular type of bright lines in their spectra. She also identified 59 new gaseous nebulae, including that which came to be known as the Horsehead, familiar to many with even a passing interest in Astronomy.  She discovered 94 Wolf-Rayet stars and, along with Edward Pickering and Henry Norris Russell, Fleming is credited with the discovery of white dwarf-type stars. Additionally, she found ``ninety-one stars of the Fifth Type, Class O, and sixty-nine stars of the Orion type having bright hydrogen lines''  \citep{Cannon_1911}. In 1906 -- in recognition of her contributions to the study of Astronomy -- Fleming became the first American woman to be named an Honorary Member of the Royal Astronomical Society of London. She was also appointed as an Honorary Fellow in Astronomy at Wellesley College. Also, in 1906, Fleming was elected as an Honorary Member of the Sociedad Astr\'onomica de Mexico and received its Guadelupe Almendaro Gold Medal for the discovery of new stars. 
    
    Early in May, 1911, Fleming was feeling ``not so well as usual'' and entered a hospital to rest \citep{Cannon_1911}. It turned out that her extreme fatigue was due to a severe case of pneumonia which proved to be fatal. She passed away on May 21, 1911.  In the Sixty-Sixth Annual Report of the Director of the Astronomical Observatory of Harvard College which was published following Fleming's death, Edward Pickering wrote of the Observatory's ``severe loss'' \citep{Pickering_1912}:
\begin{quote}
Mrs. Fleming's record as a discoverer of new stars, of stars of the fifth type, and of other objects having peculiar spectra, was unequaled. Her gifts as an administrative officer, especially in the preparation of the \emph{Annnals}, although seriously interfering with her scientific work, were of the greatest value to the Observatory.
\end{quote}
    
    On Monday morning, May 22, 1911, \emph{The Boston Herald} featured the passing of Fleming on its front page with the headline ``Leading Woman of Science Dies in a Hospital.'' Of Williamina and her work a subheadline stated that ``In 20 years of work at the little brick building on Observatory Hill in North Cambridge … she made more discoveries than all other astronomers had made in 200 years'' \citep{Herald_1911}.
     
     Writing on behalf of the Royal Astronomical Society in its February, 1912, issue of the \emph{Monthly Notices}, its editor Herbert H. Turner paid the following tribute to Fleming \citep{Turner_1912}:
\begin{quote}
As an astronomer Mrs. Fleming was somewhat exceptional in being a woman; and in putting her work alongside that of others, it would be unjust not to remember that she left her heavy daily labours at the observatory to undertake on her return home those household cares of which a man usually expects to be relieved. She was fully equal to the double task, as those who have had the good fortune to be her guests can testify; and it is perhaps worthy of record, as indicating how lightly the double burden sat on her, that she yielded to none in her enjoyment of a football match, especially a match between Harvard and Yale.
\end{quote}
             
             As the years have passed, Fleming's name has been associated with the grouping of the Harvard Observatory's Women employees and she is often referred to as one of a number of ``Computers.'' But, in her many years at the HCO in which she drove forward the work of the Draper Memorial and discovered, analyzed, and classified innumerable stars of various types, she considered herself to be an astronomer.  On May 10, 1907, Williamina Fleming submitted to the Massachusetts District Court her Petition for Naturalization requesting ``to be admitted a citizen of the United States'' which was signed and witnessed by Director Edward C. Pickering and Harvard Astronomy Professor Solon I. Bailey. On the document, Fleming listed her occupation as ``Astronomer'' \citep{Naturalization}.
    
    For the final word on the preferred nomenclature for Fleming's occupation, we may look to her gravestone in the Mount Auburn Cemetery in Cambridge, Massachusetts, which reads, simply: Williamina Paton Fleming, Astronomer. 
    
    In 1970, the International Astronomical Union named a Lunar crater \citep{FlemingLunarCrater} and in 1991 Asteroid 5747 \citep{WilliaminaAsteroid} after her.

	\section{Stellar Spectroscopy at the Harvard College Observatory}
	\label{Stellar}

	\subsection{Stellar Photometry}
	\label{Photometry}
		
	Early in his directorship of the Harvard College Observatory, which began in 1877, Edward Charles Pickering took note of research programs at other institutions and found the field of photometry to be lacking as a major undertaking. On October 25, 1879, Pickering launched a project in which the brightness of stars as recorded on photographic glass plates would be accurately measured \citep{Bailey_1931}, p. 188. He hoped to create a uniform scale of stellar magnitude and, with the assistance of the well-known Cambridgeport, Massachusetts, astronomical optics firm Alvan Clark \& Sons, he would go about creating the specialized instrumentation required to do this work \citep{Plotkin_1990}. Of Pickering’s many photometric inventions, the Meridian Photometer is one of the best known and became the project's most relied-upon instrument. It made use of Harvard's Great Refractor\footnote{The acquisition of this 15-inch telescope by Harvard in 1847 was funded by the Boston citizenry whose interest in astronomy was aroused by the Great Comet of 1843 \citep{Bailey_1931}, p. 84.} and dual-objective optics that enabled comparing the brightness of a given star with that of Polaris chosen as a standard \citep{Lewis_1885}. In 1882, the Observatory published its first work devoted to these photometric efforts, known as the \emph{Harvard Photometry}, which appeared in Volume XIV of the \emph{Annals of the Astronomical Observatory of Harvard College}. The work included analysis of the magnitudes of 4,260 stars brighter than 6th magnitude and received a stellar review which appeared in 1885 in Volume 8 of the publication \emph{Observatory}, a monthly review of astronomy \citep{Lewis_1885}: 
	\begin{quote}
To say that the volume recently issued from the Observatory of Harvard College is, in this branch of astronomy, epoch making, is to do no more than justice to a work which must henceforth be regarded as at once the foundation and treasury of scientific stellar photometry.
\end{quote}
Over the course of three years, Pickering and the HCO staff members who had worked on the \emph{Photometry} had managed to assign a designation of magnitude to every star visible from Cambridge, Massachusetts \citep{Sobel_2016}, p. 12. 

During this time, besides fulfilling her copying duties, Williamina Fleming, see Subsection \ref{Fleming}, learned how to measure the brightness of stars on photographic plates and to establish their magnitudes. She also became familiar with the appropriate methods of handling these photographs on glass and placing them in the special frame, backlit by a mirror, which allowed her to view the images. The process also involved the inspection of the glass plates in order to determine the quality of each image. The preservation of these fragile images, eventually numbering 200,000 during her tenure, along with their proper cataloging and overall care, fell directly on Williamina's shoulders as, in 1899, she would be granted the title \emph{Curator of Astronomical Photographs} by the Harvard Corporation, thus becoming the first woman to hold a professional position at Harvard.   
	
\subsection{Mapping Out Stars According to their Spectra: The Henry Draper Memorial Catalog}
\label{Mapping}
	
An astronomer and pioneering astrophotographer by avocation if not profession, Henry Draper had passed away merely days after a dinner party at which the HCO Director was in attendance along with many members of the National Academy of Sciences. Draper, thanks in part to the largesse and enthusiasm of his wife, Anna Palmer Draper (1839-1914), had followed in the path of his own father, John William Draper (1811-1882),\footnote{John William Draper was the first president of the American Chemical Society.} in the newly-developed pursuit of astrophotography. Supported by family wealth, Henry Draper was able to retire from his position as Professor of Physiology at New York University in 1882 -- abandoning academia in favor of pursuing his free-lancing interests in astronomy. 
 
   It was not only photography of the stars and other astronomical objects that had inspired Henry Draper’s imagination, but the much more specialized and challenging process of imaging their spectra that captivated him. Capturing the spectrum of a particular star was more technically difficult than merely imaging the object itself and the process of trial and error was heavily involved in this undertaking.\footnote{His first success occurred in 1872 when he photographed the spectrum of the star Vega or, alpha Lyrae. Among his many triumphs that followed was the imaging of the spectra of alpha Aquila, Arcturus, Capella, the Moon, Venus, Mars, and Jupiter \citep{Barker_1888} and that of the Great Nebula of Orion \citep{NYT_1888}.}  By the time of his untimely death at the age of forty-five, Draper had taken more than one hundred spectral images. 
   
    In early 1883, Pickering sent a letter of condolence to Anna Draper, in which he suggested that the HCO might, with her blessing, continue the spectroscopic work of her late husband.  Pickering, although aware that astronomical ``firsts'' were especially notable, knew that if others achieved success in spectral astrophotography, the efforts of Henry Draper would become a mere footnote.  Mrs. Draper was pleased to hear from Pickering and, after some deliberation, decided to fund his idea which, she perceived, would be a suitable tribute to her late husband. Originally, she had had the idea to establish an observatory in her husband's name and to run it on her own, but that plan never came to fruition. Thus, she agreed with Pickering that February 14, 1886 would stand as the official date of the establishment of the great undertaking by the HCO, named the Henry Draper Memorial \citep{Zaban_1971}, p. 228. And, not only stars of the Northern Hemisphere were to be included in the spectroscopic survey but also those of the Southern skies. Fortunately for Pickering and the fate of the enterprise, the will of engineer Uriah A. Boyden was finally settled in 1887 following five years of legal deliberations which resulted in a sum of money being left to Harvard for the sole purpose of establishing a high-altitude observatory. This funding helped to create the Boyden Department as the Harvard College Observatory’s Southern Hemisphere site came to be known \citep{Zaban_1971}, pp. 246-252.
 	
	      With a steady source of financial backing, thanks mainly to Mrs. Draper’s generosity, Pickering was able to provide both the northern and southern departments of the Observatory with the equipment needed. Mrs. Draper transferred her husband's 11-inch-refractor to Pickering on loan for however long it would take to complete the required work. Another telescope was purchased in 1885 with money granted by Alexander Dallas Bache, a physicist and the former Superintendent of the United States Coast Survey. The Bache fund was administered by the National Academy of Sciences. The 8-inch Bache telescope was eventually dismantled and sent from Cambridge, Massachusetts, to Arequipa, Peru. The instrument had been very productive at the Observatory in Cambridge, so Mrs. Draper ordered a new 8-inch telescope that would be used there instead of the Bache \citep{Hoffleit_1991}. Several other instruments from the Drapers' collection, along with other assorted apparatus, including lenses and various prisms, were given over to the work of the Draper Memorial or purchased as needed.
	      
   In 1888, Pickering issued a further appeal for donations with the aim of implementing his astrophotographic agenda for the inventory and mapping of the heavens. This resulted in a gift which, as Pickering had hoped, provided a larger telescope and lens for the project. Catherine Wolfe Bruce, a philanthropist who happened to be a devotee of astronomical science, had received a sizable inheritance from her father. She agreed to provide \$50,000 toward building a 24-inch photographic telescope. Miss Bruce was so enthusiastic about funding astronomical endeavors, and Pickering was so persuasive in convincing her of the great need for support of astronomical institutions in general, that she agreed to participate in worthy efforts at observatories worldwide. The two accepted applications for funding and chose ``five scientists in the United States to receive her support'' along with ten other grants to astronomers working in England, Norway, Russia, India, and Africa \citep{Sobel_2016}, p. 48. Following installation and testing in Cambridge beginning in 1893, the Bruce telescope was conveyed to Arequipa in 1896. Pickering himself donated \$100,000 of his own money anonymously to further assist the project. Additional income was derived from the bequest of Robert Treat Paine, grandson of one of the original signers of the Declaration of Independence, who was a lawyer and an amateur astronomer and who left \$250,000 to the Harvard College Observatory. 
      
    Late in 1888,  HCO staff member Solon Bailey departed for Peru on a steamship and railway journey, finally reaching Lima in March 1889. From there, he traveled to the town of Chosica to explore the local environs and inspected various mountain peaks in search of a suitable site to establish Harvard's southern hemisphere observatory. An unnamed peak was finally chosen high above the town of Chosica and named ``Mount Harvard'' \citep{Zaban_1971}, p. 291. Bailey's first impulse had been to name the peak ``Mount Pickering'' but, when informed by mail, the Director declined the honor.  The first necessity was to build a trail by which to deliver the nearly hundred boxes along with the 8-inch Bache Telescope up the mountain. In the interest of hurrying the process along, Bailey and his family helped to provide labor. Of the area in general, Bailey noted \citep{Bailey_1895}:
    \begin{quote}
    From the time of our first visit to Mount Harvard on March 11 throughout the rest of the year the region seemed extremely dry and barren. The ravine, however, through which our path to the mountain led, showed unmistakable evidence of tremendous rainfall sometime in the past.
    \end{quote}
Once the observatory was finally established and in working order, the process of imaging the southern sky and recording its splendors on astrophotographic glass plates commenced. The first boxes arrived back at the Harvard College Observatory in Cambridge, Massachusetts, in August 1889, though it was found that some of the plates had suffered slight damage during shipment. Photographic work continued at Mount Harvard through the Peruvian autumn. Meanwhile, other potentially desirable sites for a more permanent establishment were inspected including the Atacama Desert plateau in Chile.  Eventually, in late 1889 and early 1890, fog and mists enveloped Mount Harvard and were followed by heavy rains. Bailey had reported to Pickering that an alternative site might exist at Arequipa, Peru. As the year progressed, the weather remained unstable through the summer with more heavy rain. The buildings at the Observatory site, with walls constructed of building paper, began to sag and collapse. Given the discouraging reports from Peru, Edward Pickering had decided to move the Boyden Station to Arequipa as its permanent site, and work ceased at Mount Harvard. The original site was dismantled and the instrumentation packed for shipment. Bailey and his party moved on to Arequipa where the observational and photographic work continued until May 9, 1891. Over the course of two years, Solon Bailey had managed to establish the magnitudes of approximately 8,000 bright stars thus completing the \emph{Harvard Photometry} \citep{Zaban_1971}, p. 291. He had also obtained around 2,500 photographic plates of the spectra of the southern stars with the aid of his brother Marshall, a professional photographer, and of his Peruvian assistant Elias Vieyra. On May 15, 1891, Solon Bailey and his family departed Arequipa. After an unfortunate interregnum at Arequipa, see Subsection \ref{Pickering}, Solon Bailey returned there on February 25, 1893, and remained in charge of the outpost until 1919.
       
\begin{figure}
		\centering
		\includegraphics[scale=0.51]{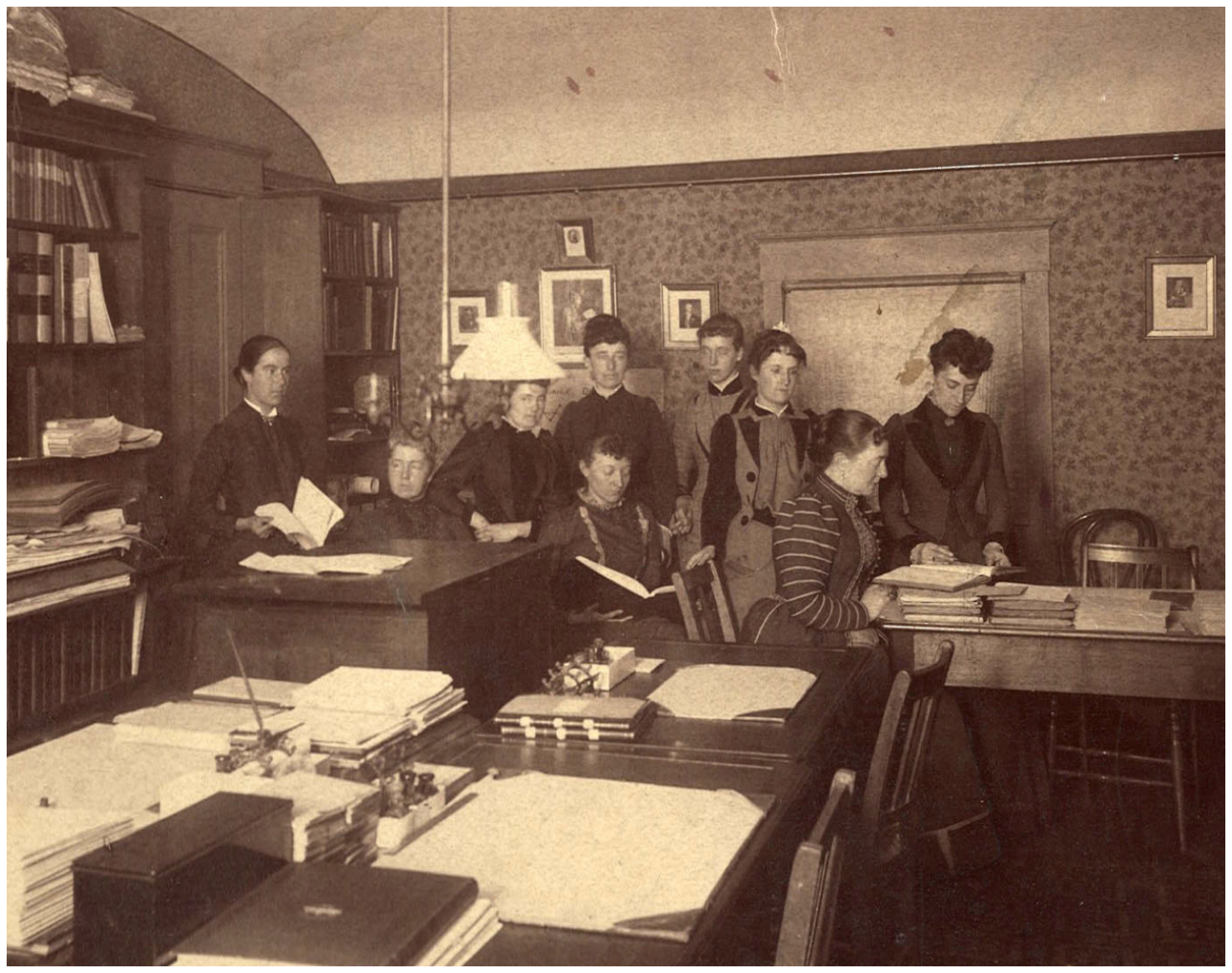} 
		\caption{\label{fig:PhotoMsDraperVisit} Mary Anna Draper (1838-1914), seated at right, during her visit to the HCO in 1891. She sponsored the Henry Draper Catalog named so in honor of her late husband, the physician and amateur astronomer -- and pioneer of astrophotography -- Henry Draper (1837-1882). Standing next to her at the table is Williamina Fleming, the Curator of the Astronomical Photographic Glass Plate Collection and supervisor of the women astronomers at the HCO.  HUV 1210 (9-5), olvwork289692. Harvard University Archives, 1891.} 
	\end{figure}

In Williamina Fleming, Pickering had found the most capable person possible to keep the work of his Observatory on track. Besides her scientific work -- the reduction of the data contained in the astrophotographic images -- Fleming, with her great diligence, determination, and attention to detail, became the foreperson who was put in charge of the Women Astronomical Computers, their hiring, workflow, and output. She also oversaw the entire process of publishing all Observatory-related materials. Fig. \ref{fig:PhotoMsDraperVisit} shows HCO's Women Astronomical Computers during a visit by Anna Draper.
 
\begin{figure}
		\centering
		\includegraphics[scale=0.5]{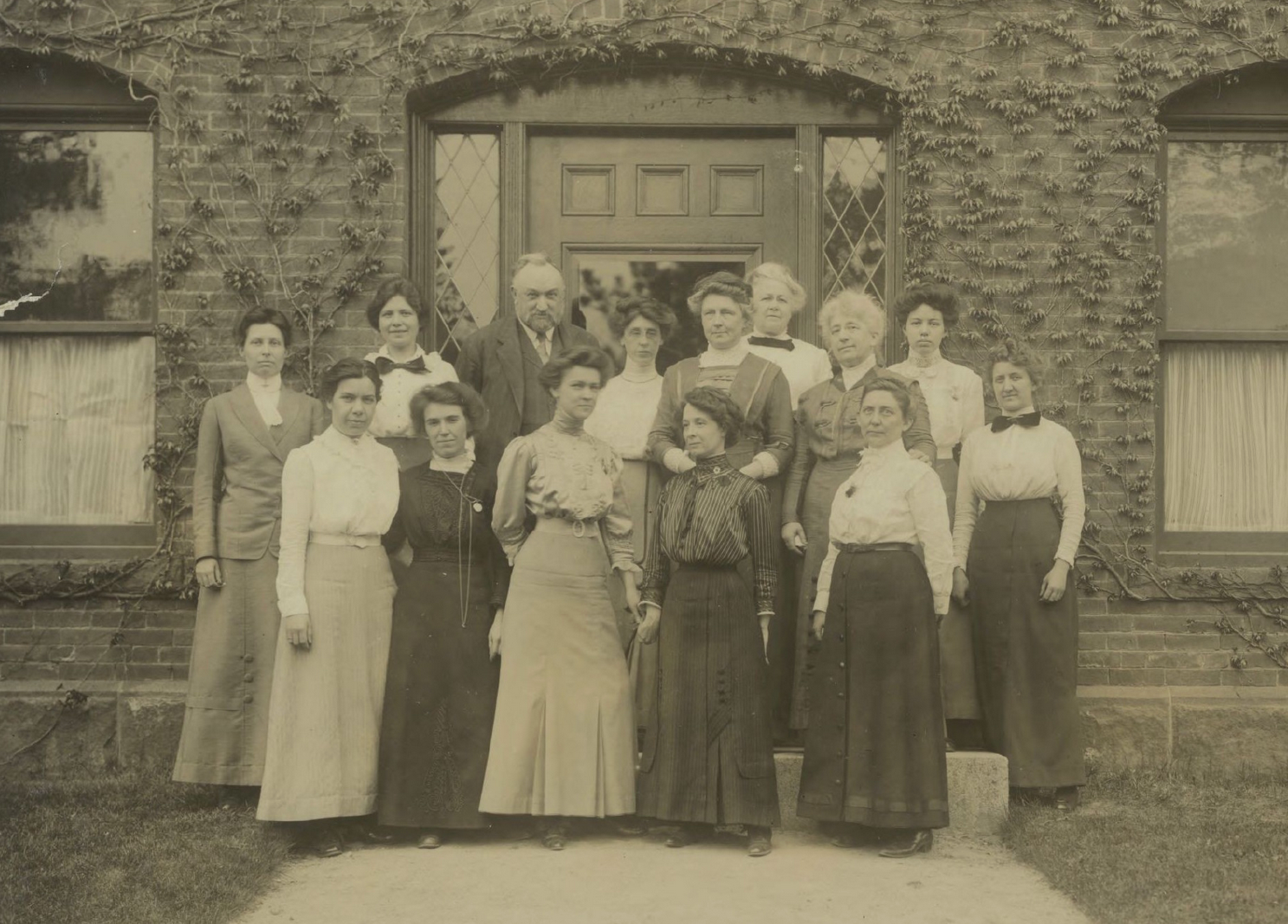} 
		\caption{\label{fig:PhotoPickering_sHarem} \small{Female members of Pickering's team, also known as ``Pickering's Harem'' \citep{Sobel_2016}, pp. 105-122. At the far left is Margaret Harwood (A.B. Radcliffe 1907, M.A. University of California 1916), who was later ap­pointed director of the Maria Mitchell Observatory. Beside her in the back row is Mollie O'Reilly, a computer from 1906 to 1918. Next to Pickering is Edith Gill, a computer since 1889. Then Annie Jump Cannon (B.A. Wellesley 1884), who at this time was about halfway through classifying stellar spectra for the Draper Catalogue \citep{Draper}. Behind Miss Cannon is Evelyn Leland, a computer from 1889 to 1925. Next is Florence Cushman, a computer since 1888. Behind Miss Cushman is Marion Whyte, who worked for Miss Cannon as a recorder from 1911 to 1913. At the far right of this row is Grace Brooks, a computer from 1906 to 1920. Ahead of Miss Harwood in the front row is Arville Walker (A.B. Radcliffe 1906), who served as assistant from 1906 until 1922 and later secretary to Pickering's successor Harlow Shapley. The next woman may be Johanna Mackie, an assistant from 1903 to 1920. In front of Pickering is Alta Carpenter, a computer from 1906 to 1920. Next is Mabel Gill, a computer since 1892. And finally, Ida Woods (B.A. Wellesley 1893), who joined the corps of Women Computers just after graduation. In 1920 she received the first A.A.V.S.O. nova medal; by 1927, she had seven bars on it for her discoveries of novae on photographs of the Milky Way \citep{Welther_1982}. HUPSF Observatory (14), olvwork360662. Harvard University Archives, ca 1910.} } 
	\end{figure}

        By 1893, there were a total of 40 assistants working at the HCO, 17 of them women \citep{Sobel_2016}, p. 53. Fig. \ref{fig:PhotoPickering_sHarem} shows a 1913 photo of 13 female members of Pickering's team. 
        
 In a satire ``HCO Pinafore'' written in 1879 by Winslow Upton, then a telescope assistant and later professor of Astronomy at Brown University \citep{Bailey_1931}, to the tune of Gilbert \& Sullivan's 1878 comic opera ``HMS Pinafore,'' the chorus of Women Computers sang \citep{100CPG_2001}, p. 161:
 \begin{quote}
 We work from morn 'till night,\\
 For computing is our duty;\\
 We're faithful and polite,\\
 And our record book's a beauty;\\
 With Crelle and Gauss, Chauvenet and Peirce,\\
 We labor hard all day;\\
 We add, subtract, multiply and divide,\\
 And we never have time to play.\\
 \end{quote}
This telling satire was performed for the first time only in 1929, at the residence of the then HCO Director Harlow Shapley (1885-1972) for about 100 members of the American Astronomical Society. We note that when the HCO Pinafore was written, there were four women and several men computers employed by the observatory \citep{LaFortune_2001}. The satire's sting had apparently a more general aim than the HCO, with the Naval Laboratory and other government services being among its targets as well \citep{Grier_2013}, pp. 83-84. The piece also spoofed, or perhaps not \citep{Lewis_1885} and \citep{Fiss_2023}, Pickering for his choice of Polaris, a variable, as a luminosity standard (for whose variation the photometric measurements had to be corrected):
\begin{quote}
 Pole Star, to thee I sing\\
            Bright pivot of the heavens,\\
         Why are all our magnitudes\\
         Either at sixes or at sevens?\\
            I have lived hitherto\\
         Free from the breath of slander.\\
            Beloved all my crew\\
         A really popular commander.\\
         But now my prisms all rebel\\
            And ruin the photometer,\\
            And damage also, sad to tell,\\
            My fame as an astronomer.\\
            Pole Star, to thee I sing\\
            Bright pivot of the heavens,\\
         Why are all our magnitudes\\
         Either at sixes or at sevens?\\
\end{quote}

       	\subsection{Harvard Stellar Classification System}
	\label{Classification}
   
   In 1882, Pickering laid out a plan suggesting that interested parties might search for five types of stars with varying magnitudes  \citep{Pickering_1882}, p. 5.:
   \begin{quote}
    I. temporary stars, or those that shine out suddenly, sometimes with great brilliancy, and gradually fade away, otherwise known as novae; II. long-period variables, or those undergoing great variations of light; III. stars undergoing slight changes according to laws as yet unknown; IV. short period variables, or stars whose light is continually varying, but the changes are repeated with great regularity in a period not exceeding a few days; V. Algol stars, or stars which…every few days suffer a remarkable diminution in light for a few hours.
   \end{quote}    
   Toward the conclusion of his tract, Pickering provided motivation for undertaking the study of variable stars in case any reader should need convincing  \citep{Pickering_1882}, p. 14-15:
     \begin{quote}
     Apart from the value of the results attained it is believed that many amateurs will find it a benefit to accustom themselves to work in a systematic manner, and that they will thus receive a training in their work not otherwise easily obtained outside of a large observatory. The lesson should be taught that time spent at a telescope is nearly wasted unless results are secured worthy of publication and having a permanent value.
     \end{quote} 

Heeding Edward Pickering's clarion call, many amateurs followed through with the plan of observing variable stars and reported the resulting data to the HCO. Thus Williamina Fleming scrutinized not only astrophotographic plates taken in Cambridge, Massachusetts, and those shipped from Arequipa, but she also sought to verify claims by the amateurs of variability in regard to particular stars. In her thorough investigations of these stars, she discovered that a preponderance of those categorized as variables were Type III stars according to  Angelo Secchi's system based on spectral appearance. Initially, Secchi's system contained two classes, one of which consisted of red and yellow stars like the sun while the other contained blue stars such as Sirius \citep{Holberg_2007}, p. 87.  As he was exposed to more and more spectra, Secchi changed his classifications to allow for five different types: Class I: blue stars such as Sirius which display a small number of hydrogen absorption lines; Class II: stars similar to the sun, with spectra rich in narrow dark lines; Class III: reddish stars such as Betelgeuse which display broad dark bands; Class IV: stars with strong carbon lines \citep{Holberg_2007}, p. 87. In 1877, Secchi added a fifth class of emission-line stars. Thus a natural part of Williamina Fleming's job was devising a new classification system \citep{AJC_1915}: 
\begin{quote}
The divisions into five types made by Secchi proved altogether inadequate to represent the numerous differences seen on the photographs. A new system had to be adopted which would permit the reader to understand the various aspects of the spectra as shown by the photographs. …This classification is purely empirical, being based wholly on the external appearances, without any idea of expressing differences in temperature or stages of evolution.
\end{quote} 
Fleming elaborated on Secchi's spectral classes and added a number of subdivisions, each of which received a letter designation: A, B, C, D, E, F, G, H, I, K, L, M  and N. Class Q would be added later for stars with spectra that didn’t really fit into any other category. In Fleming's system, 
\begin{quote}
Stars of Type A and B were bright blue stars like Sirius and those in Orion…while stars of Type F and G had spectra similar to that of the sun, and K and M stars were characterized by reddish spectra containing strong dark bands, in addition to the narrow metallic lines seen in other stars.
\end{quote} 
The letter ``O'' was added to the classification to indicate those stars which Fleming herself considered to be ``the most striking class of stellar spectra, in fact the only one that is at once detected visually with a small telescope'' \citep{Fleming_1912}, p. 177. Included in this class would be Wolf-Rayet-type stars along with 21 stars of this spectral type that had been discovered in the Large Magellanic Cloud. A number of others were found to ``occur near the central line of the Milky Way.'' This group of stars would become known as being of the Fifth Type, of which Williamina Fleming discovered many. 

The Second Conference of Astronomers and Astrophysicists was held at Harvard in August 1898, where it was resolved that the American Society of Astrophysicists, now known as the American Astronomical Society, would come into existence. Fleming submitted a paper to the conference entitled, ``Stars of the Fifth Type in the Magellanic Clouds,'' which was read by  Edward Pickering. Up to that time, according to the paper, 92 such stars exhibiting the particular bright line spectra had been discovered. At the end of his reading of the paper, Pickering took the liberty of informing the crowd that, of those 92 stars, 78 had been discovered by Williamina Fleming (who had not mentioned this fact in the paper) and who was in attendance. A spontaneous burst of applause impelled her to come forward whereupon she responded to questions that had been generated by her paper from the crowd of astronomers in attendance \citep{Donaghe_1898}.

    Annie Jump Cannon simplified Fleming's system, combining certain classes and doing away with others, such as little-used C, D, E, H, I, L, N, and Q.  She noted that \citep{AJC_1915}, pp. 209-210:
    \begin{quote}
    A modification of the system of letters used for the Draper Catalogue was adopted [in 1897] …  The stellar sequence was found to be in some respects less complex than was at first supposed. The appearances for which some of the letters, such as C, D and E, had been assigned, were not confirmed by later and better photographs. Therefore, the letters were dropped from the sequence.
    \end{quote}
     Cannon reordered the spectral types to the now-familiar types O, B, A, F, G, K, M. The new ordering system allowed for differentiations in spectra to be expressed within the various types either by a supplementary letter such as Oa (``A broad bright band whose centre is at wavelength 4633 is the most conspicuous feature of this spectrum \citep{AJC_1912Class}, p. 66) or by the addition of a digit such as F2 (``This spectrum resembles Class F, except that there is a slight appearance of continuity in band G'' \citep{AJC_1912Class}, p. 68). Cannon's reordering aligns with the abscissa of the Hertzsprung-Russell Diagram of stellar luminosity versus temperature. 
     
     In 1910, Edward Pickering attended the fourth meeting of the International Union for Cooperation in Solar Research, which was held on the summit of Mount Wilson, California, at the solar observatory of the Carnegie Institution. At this time, international astronomy did not abide by a single stellar classification system. Pickering saw an opportunity for the widespread adoption of the Harvard Pickering/Fleming/Cannon system. Recording his thoughts in a diary on the long railroad trip to California, he wrote that the Harvard system
\begin{quote}
will probably be the system of the world. I am repaid for my journey of 2,000 miles, had I done nothing else … My part in this will be regarded as one of the most important things I have ever done. 
\end{quote}  
   To his great relief and delight the Union gave it ``the strongest endorsement I could have desired'' \citep{Plotkin_1990}, p. 57. 
   
   \subsection{Variable Stars}
   \label{Variable}
   
   Formerly, verification of a variable -- a star with a time-dependent luminosity -- required a tedious process of following the star over a series of plates, taken at various times, and determining whether its magnitude actually did vary and, if so, whether there was a regular, discernible pattern of brightening and darkening. In 1881, Edward Pickering forwarded an announcement to the journal \emph{Astronomische Nachrichten} entitled ``New Variable Star in Puppis'' \citep{Pickering_1881}. The star, in Pickering’s opinion, was a Mira-type star as it resembled Omicron Ceti, an old, pulsating red giant known as Mira or ``The Wonderful'' in Latin. Mira's variability had been recorded by German pastor and amateur astronomer David Fabricius (1564-1617) beginning on August 3, 1596, but evidence of its capricious nature is hinted at in earlier Chinese, Babylonian, and Greek records.  In his account, Pickering went on to write that:
   \begin{quote}
   Its variability was first suspected from the character of its spectrum, which resembled that of some other variable stars of long period.  A method of detecting such objects is thus indicated, far more rapid than the usual one of repeatedly observing the magnitude of the same star.
   \end{quote}
   Williamina Fleming proved the Director's statement as, she herself, went on to discover 125 Mira-type variables spectroscopically \citep{Hoffleit_1997}. In 1907, she looked back upon this newer, more convenient method of determining variables \citep{Fleming_1907Variable}:
   \begin{quote}
     The examination of photographs of the Henry Draper Memorial has led to the discovery of a large number of variable stars of long period. The greater portion of these has been found from the characteristic spectrum, Md, which indicates the third type traversed by bright hydrogen lines. While many variable stars of long period have spectra of the fourth type, and a few have spectra of the third type in which the hydrogen lines are not bright in the photographs, no case has yet been found in which a star having the spectrum Md, described above, is not a variable.
     \end{quote}
   Among the HCO staff members -- north and south -- there was a competitive spirit involved in the finding of variable stars. For instance, Solon Bailey was interested in investigating variables found in globular clusters. Unfortunately for Fleming, her finding that particular formations of spectra were indicators of a star's variability provided a key to their discovery. One no longer needed to consecutively view the star itself in order to determine whether it was variable but had only to see its spectrum as viewed on a glass plate. Soon, plates began to arrive from Arequipa with the variable stars already marked. The assistants at the Boyden Station who were responsible for imaging the southern skies, with the encouragement of Bailey, would often inspect the plates following exposure and look for the variables thus partaking in the thrill of discovery and a modicum of credit for the achievement. As a result, Fleming would receive plates back in Cambridge that had already been marked, and she was quite vexed. She felt that, since the plates were her responsibility, the process of ferreting out the variables should be hers as well. Very much displeased, she made her feelings known to Director Pickering.  A message was dispatched to Bailey explaining that Mrs. Fleming's work was being duplicated and that the Peruvian assistants would, in the current situation, receive credit for the discoveries that they made while it was she who was still involved in doing the greater part of the work. As Pickering wrote on September 29, 1897 \citep{Zaban_1971}, pp. 354-355: 
\begin{quote}
   She is obliged to measure the positions, the variations in brightness, if any, and to identify the individual lines, classify the object and see if it is a catalogue star. She also has to reexamine the plates since the fainter objects, including about half of the peculiar objects, and as many more having slight peculiarities are omitted. All of this is part of her regular routine work and has been for the past ten years, and much of it could not be done in Peru …The delay might also prevent the early discovery of a new star or other object of special interest.
\end{quote}
Pickering hoped that Bailey might be able to provide some means of mitigating this ``source of friction'' but, instead, Bailey came to the defense of his assistants. He didn't think that it was very realistic, or very fair, to prevent the employees at Arequipa from looking over the photographic results and marking any unusual objects. He added: 
\begin{quote}
Personally, I think that the ability to make first class plates is greater than that required in the mere picking up of new objects by certain well known characteristics, as it has been done generally by the assistants here. But as the latter has been publicly recognized and the former seldom or never, it is perhaps not strange that an ambitious assistant should desire to try also the latter… Mrs. Fleming is not the only one who has felt vexed at times.
\end{quote}
   Despite the occasional difficulties, personal and otherwise, and misunderstandings that sometimes arose from the establishment and maintenance of the Arequipa station, the photographic results produced there were more than suitable as Pickering would later express with great satisfaction \citep{Pickering_1897Southern}:
\begin{quote}
A distinguishing feature of the climate of Arequipa is the great steadiness of the air. The value of this location as an observing station is largely due to this fact. Good definition, under high powers, is obtained there on many more nights than in Europe, or in the United States, where nine-tenths of the observatories of the world are presently located.
\end{quote}
 
      Already in 1881,  Edward Pickering submitted an advisory from Cambridge, Massachusetts to \emph{Astronomische Nachrichten}, which began \citep{Pickering_Schreiben_1881}
      \begin{quote}
      Herewith I send you a list of objects discovered here, and having some peculiarity either of color or of the distribution of light in their spectra. 
      \end{quote}
      The list consisted of thirty-nine stars, many of which were provided with a corresponding remark such as ``Perhaps variable; red;'' ``Bands distinct; fine specimen,'' or the name of its discoverer. This list was continued in the \emph{Astronomische Nachrichten} the following year as ``Stars with peculiar spectra, discovered at the Astronomical Observatory of Harvard College'' \citep{Pickering_1882Stars}. This series of announcements of the discovery of stars with peculiar spectra at the HCO would continue on for several decades and see publication in HCO \emph{Circulars}, the journal \emph{Sidereal Messenger} (renamed in 1892  \emph{Astronomy and Astro-Physics}) and, also, in 1917, the finding of a ``Peculiar Asteroid'' was announced in an issue of the \emph{Harvard College Observatory Bulletin}.  Through the years, quite often, a single announcement would appear in several different titles simultaneously -- verbatim except for acknowledgment of the particular publication's title. 
       
   On September 13, 1890, another notice appeared in the \emph{Sidereal Messenger}; while the notice had been ``Communicated by Edward C. Pickering, Director of Harvard College Observatory,'' the submission was credited to ``M. Fleming'' \citep{Pick_Flem_1890}.  In the notice related to plates taken at Chosica, Peru, Fleming described the spectral peculiarities of stars and the suspicion of their variability due to the bright hydrogen lines exhibited photographically. Her suspicions were ultimately proven correct. From this publication onward, notices of spectrally peculiar stars would either be submitted by Fleming or, when authored by Director Pickering, they would mainly feature her findings. A year later, in 1891, another submission to \emph{Astronomische Nachrichten} by Fleming, and, once again, ``communicated'' by Pickering, announced the finding of a star showing ``a spectrum consisting mainly of bright lines, and similar to that of the Wolff and Rayet stars in Cygnus'' \citep{Pick_Flem_1891}. Fleming would go on to discover 94 of the 107 known stars of the Wolf-Rayet type. 
   
   In 1912, Edward Pickering wrote \citep{Pickering_1915}:
   \begin{quote}
   On one plate more than a thousand spectra were classified. The late Williamina P. Fleming, Curator of Astronomical Photographs, from an examination of these plates, discovered several thousand objects having peculiar spectra. In fact, probably few bright objects of this class escaped her. Of the nineteen new stars, known to have appeared during the progress of this work, she discovered ten, and five others were found by other observers here. In this work alone also, the number of stars of the peculiar class known as fifth type, has been increased from seventeen to one hundred and eight.
   \end{quote} 
   Any star detected by Mrs. Fleming that did not fit into Harvard's established classes was considered to be \emph{peculiar}.  The reason might be due to some variation in the spectrum such as an unexpectedly thick or thin spectral band or an unusually bright, dark, or curiously placed spectral line. Many such stars were of a type that had not yet been closely studied or were undergoing hitherto unseen processes in their evolution. Certain elements, the existence of which altered the appearance of particular spectra were, as of yet, undetected when Fleming was classifying her multitudes of stars. Neither were other processes obvious such as that of ionization which strips away electrons due to elemental superheating and changes the elements' spectra.
   
  Even Henry Norris Russell, the acknowledged expert on stellar composition, opined \citep{Russell_1914}:
  \begin{quote}
  The first great problem of stellar spectroscopy is the identification of this predominant cause of the spectral differences. The hypothesis which suggested itself immediately upon the first studies of stellar spectra was that the differences arose from variations in the chemical compositions of the stars.
  \end{quote}  
   It would not be until further work was carried out by Cecilia Payne-Gaposchkin \citep{Cecilia_1925}, that an understanding of stellar nucleosynthesis would begin to take hold. Payne-Gaposchkin measured the output of the energies from stars of the various stellar classes and correlated their spectra with their elemental makeup. She observed and measured the abundances of the various elements related to each class and took into consideration the effects of ionization, employing astrophysicist Meghnad Saha’s 1920 theory \citep{Hearnshaw_2014}, pp. 138-140. As summarized by William Sheehan \citep{Sheehan_2015}, p. 162:
 \begin{quote}
 The spectral sequence is thus a result of almost entirely of temperature progression in the atmosphere of stars … The higher the temperature, the faster the atoms are moving, and the more electrons will be stripped away. Since the number of atoms in one state of ionization versus another depends on competition between collisions of all kinds and radiation causing ionization and rates of electron-ion collisions producing recombination, temperature – not chemical composition – determines which absorption lines appear in a star's spectrum.
 \end{quote} 
	
	\section{Discovery of the Peculiar Spectrum of $\zeta$-Puppis}
	\label{Zeta}
	
\begin{figure}
		\centering
		\includegraphics[scale=0.6]{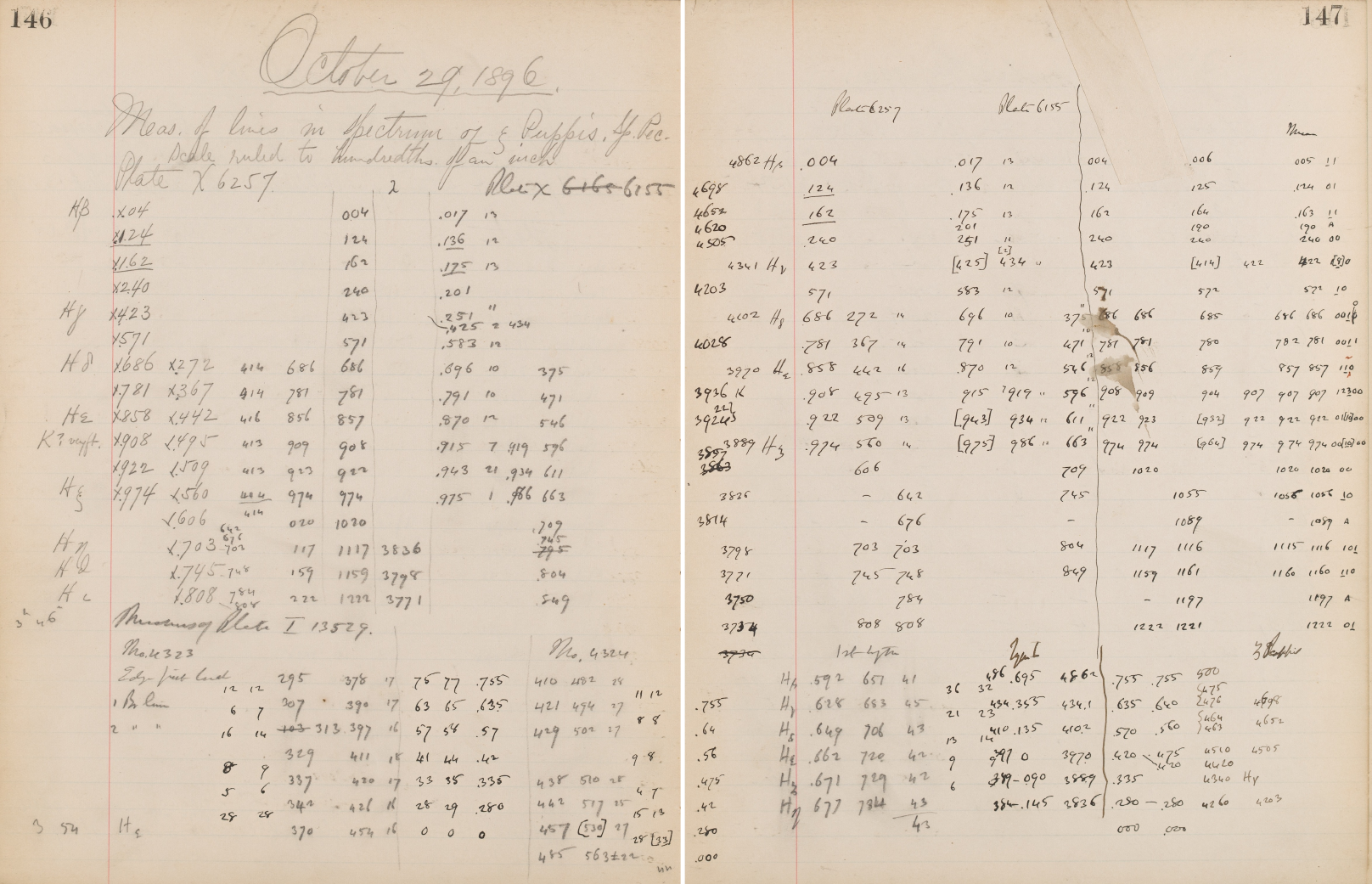} 
		\caption{\label{fig:MF_Logbook_1896} Pages 146 and 147 from Williamina Fleming's logbook of October 29, 1896 \citep{Fleming_1896_Logbook}, pertaining to the Glass Plate Number X6275 shown in Fig. \ref{fig:zeta Puppis Plate and Jacket_06257_1896}. Note that on the left margin of page 147 are listed the wavelengths of the emission lines of the Pickering series of He$^+$: 4505, 4203, 4028, 3923 (average of 3922 and 3924), 3857, and 3814 \AA. Courtesy of Harvard College Observatory.} 
	\end{figure}

\begin{figure}
		\centering
		\includegraphics[scale=0.6]{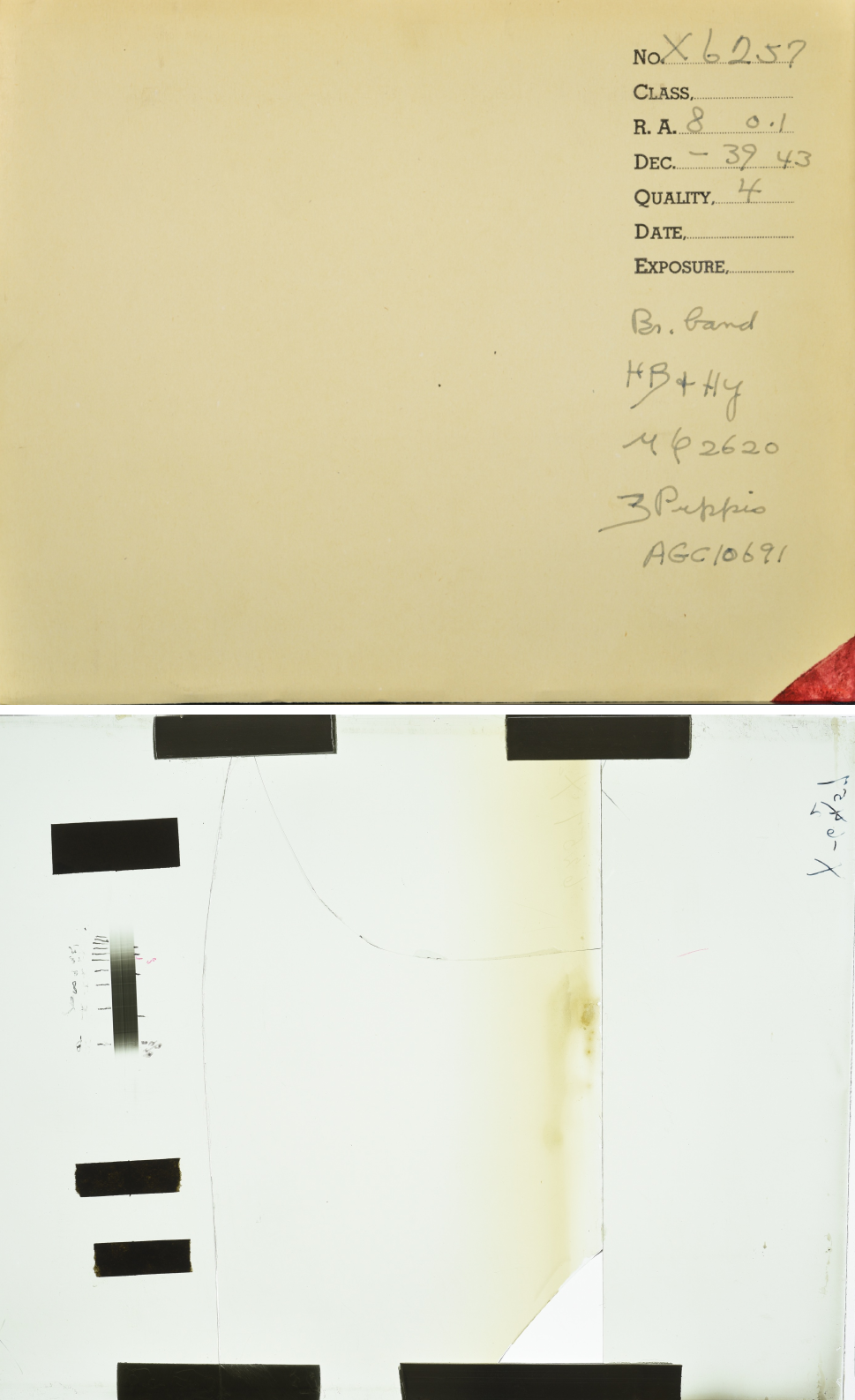} 
		\caption{\label{fig:zeta_Puppis_Plate_and_Jacket_06257_1896} Photographic glass plate number X6275  with an image of the spectrum of $\zeta$ Puppis taken on 17 December 1894 at HCO's Boyden Station in Arequipa, Peru. Also shown is a remake of the original plate jacket. Courtesy of Harvard College Observatory.} 
	\end{figure}

	   In one of Williamina Fleming's many logbooks, the notes dated October 29, 1896, stand out as particularly striking \citep{Fleming_1896_Logbook}. On pages 146-147 in Sequence No. 9 of her logbook series, see Fig. \ref{fig:MF_Logbook_1896}, in which she examined photographic plates for the presence of variable stars, appears the note ``Meas. of lines in Spectrum of $\zeta$ Puppis,'' which records her documentation of the wavelengths found in the image borne on plate X6257, see Fig. \ref{fig:zeta_Puppis_Plate_and_Jacket_06257_1896}.
	   
\begin{figure}
		\centering
		\includegraphics[scale=0.51]{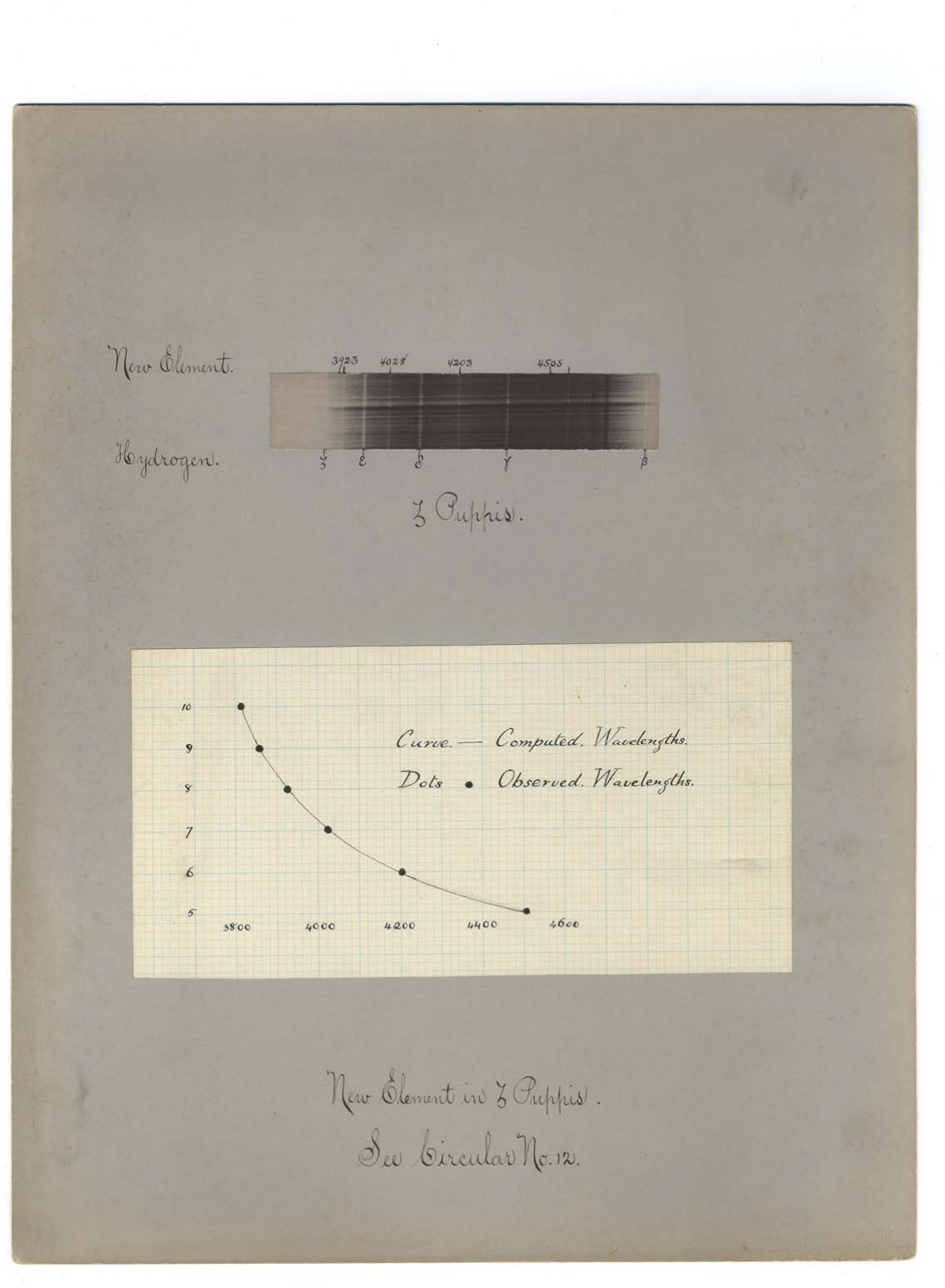} 
		\caption{(a) Upper panel: Spectrum of $\zeta$ Puppis photographed at HCO's Boyden Station in Arequipa, Peru, and evaluated by Williamina Fleming on October 29, 1896, cf. Figs. \ref{fig:zeta_Puppis_Plate_and_Jacket_06257_1896} and \ref{fig:MF_Logbook_1896}. The wavelengths shown along the upper abscissa pertain to the spectrum of the ``New Element,'' i.e., to the Pickering series of He$^+$, while the absorption lines of the Balmer series ($\beta$ through $\zeta$) of atomic hydrogen are shown on the lower abscissa and serve as wavelength standards. (b) Lower panel: A numerical fit, $m=2\sqrt{\frac{\lambda+1032}{\lambda-3618}}$, of the wavelengths $\lambda$ [\AA] of the Pickering series of He$^+$ for $m=5-10$. This fit was obtained by inverting the empirical Balmer-series-like fit, Eq. (\ref{eq:PickeringFudgedSeries}), featured in \citep{Pickering_1896}. Note that the handwriting in the panels is likely Williamina Fleming's. Published with permission from the Collection of Historical Scientific Instruments, Harvard University.} 
	\label{fig:D001012_pro} 
	\end{figure}

The plate series X denoted spectral plates with images obtained at Arequipa. With astounding speed, Williamina Fleming's documentation appeared swiftly in print in the form of HCO \emph{Circular} No. 12 on November 2, 1896 \citep{Pickering_1896Crux}. The same announcement, though slightly tailored to the individual journals, also appeared in the December 1896 issues of both \emph{Astronomische Nachrichten}\citep{Pickering_1896Nachr} and \emph{Astrophysical Journal} \citep{Pickering_1896Astr} under Pickering's name. 
In these announcements, Pickering wrote: 
\begin{quote}
A list of stars having peculiar spectra and found by Mrs. Fleming in her regular examination of the Draper Memorial photographs are given in the annexed table.
\end{quote} 
and then goes on to say that the first star in the list provided, $\zeta$ Puppis, exhibits a spectrum 
\begin{quote}
which is very remarkable and unlike any other yet obtained. The continuous spectrum is traversed by three systems of lines, First, the hydrogen lines and the line K, which are dark, as in stars of the first type. Second, two bright bands or lines whose approximate wave lengths are 4652 and 4698, which may be identical with the adjacent lines in spectra of the fifth type. Third, a series of lines whose approximate wave lengths are 3814, 3857, 3923; 4028, 4203, and 4505, the last line being very faint. \emph{These six lines form a rhythmical series like that of hydrogen and apparently are due to some element not yet found in other stars or on earth} [our emphasis].
\end{quote}
He then concluded that ``The formula of Balmer will not represent this series'' -- known since as the \emph{Pickering Series} -- but went on to show that when suitably modified, a formula involving squares of an integer pertaining to the individual lines fit the spectrum, see Fig. \ref{fig:D001012_pro},
\begin{equation}
\label{eq:PickeringFudgedSeries}
\lambda=4650\frac{m^2}{m^2-4}-1032
\end{equation}
with $\lambda$ the wavelength in \AA ~and $m$ an integer. These wavelengths were featured along with the Balmer-series-like fit, cf. Eq. (\ref{eq:PickeringFudgedSeries}), in \citep{Pickering_1896Nachr} and \citep{Pickering_1896Astr}, whose manuscripts were expedited to  \emph{Astronomische Nachrichten} and \emph{Astrophysical Journal} just four days later and published in December 1896. The first announcement of the discovery of the Pickering series appeared in Harvard College Observatory Circular No. 12 on November 2, 1896 \citep{Pickering_1896Crux}, the submission date of the paper. 

The formula for the Balmer series of atomic hydrogen\footnote{Balmer wrote his formula in the form $\lambda=3645\frac{m^2}{m^2-4}$ with the wavelength $\lambda$ in \AA ~for $m=5-10$, see \citep{Balmer_1885}.} was found empirically in 1885 by the Swiss high-school teacher Johann Jakob Balmer (1825-1898) and generalized in 1890 by Johannes Rydberg (1854-1919) for spectral series of other atoms (especially those of Groups 1-3) \citep{Rydberg_1890}.

In a follow-up paper ``The Spectrum of $\zeta$ Puppis'' published in 1897 in \emph{Circular} No. 16 \citep{Pickering_1897Circ}, in \emph{Astronomische Nachrichten} \citep{Pickering_1897Nachr}, and in the \emph{Astrophysical Journal} \citep{Pickering_1897Astr}, Pickering confirmed the wavelengths of the lines of the Pickering series but changed tack and ascribed them to atomic hydrogen. He reported a unifying empirical formula involving squares of an integer that reproduced the Balmer and Pickering series for the integer's even and odd values, respectively. This was very much in the spirit of the times when physicists were groping for the conceptual foundations of spectroscopy \citep{Robotti_1983}.

In the 1897 follow-up \citep{Pickering_1897}, Pickering noted: 
``Miss A. J. Cannon has found that the same series of lines [i.e., the Pickering series] occurs in the star 29 Canis Majoris.'' A related side note appears in Cecilia Payne-Gaposchkin's 1979 work \emph{Stars and Clusters}, in regard to both 29 Canis Majoris and  $\zeta$ Puppis \citep{CPG_1979}, p. 43: 
\begin{quote}
   It is not without significance that the known stars of very high mass are luminous blue stars: once they have left the main sequence, such stars cannot last long. The giant eclipsing variable UW Canis Majoris (29 Canis Majoris, another star of many names) and the second magnitude $\zeta$ Puppis, one of the hottest known stars, are of very high mass, at least thirty times the sun's. We do not know what the future holds for them.
\end{quote}  
   
   Today, we know that $\zeta$-Puppis  is a blue supergiant of type O4, one of the hottest and most luminous stars of the Milky Way with a surface temperature of $4.2\times 10^4$ K \citep{Schilbach_2008} where the ``element not yet found'' is especially abundant. 

Harvard College Observatory \emph{Circular}, No. 55 (1901) \citep{Pickering_1901Circ}, added further updates to the findings related to the spectral significance of $\zeta$ Puppis as more and better plates could provide sharper images. Of course, the phenomenon of these mysterious lines of ``cosmic'' hydrogen, or, ``proto-hydrogen,'' as labelled by Sir Norman Lockyer, attracted scientific attention as those interested tried to determine the precise cause of these fascinating spectral features \citep{Nature_1912}. 
   
   We note that in Williamina Fleming's final published work, ``Stars Having Peculiar Spectra,'' which appeared in the 1912 Volume LVI of the \emph{Annals of the Astrophysical Observatory of Harvard College}\citep{Fleming_1912}, the discoveries of the peculiar spectra of both $\zeta$ Puppis and 29 Canis Majoris are credited to Edward Pickering. The ``reference'' given for $\zeta$ Puppis in this work is the 1897 HCO \emph{Circular} No. 16, ``The Spectrum of $\zeta$ Puppis''  \citep{Pickering_1897Circ}, while that for 29 Canis Majoris is the 1897 \emph{Circular} No. 17, ``Stars Having Peculiar Spectra'' \citep{Pickering_1897Circ}, which were both written by Edward Pickering. Williamina Fleming's ``Stars Having Peculiar Spectra'' featured the ``Pickering Series'' and was published, with a Preface by Pickering dated October 21, 1912, more than a year after Williamina Fleming's death. By the time she had authored and overseen the production of her final publication, the notion of a ``Pickering Series'' may have gained such notoriety in scientific circles that she did not wish to detract from the renown of the Director and left things the way they were. 
    
    In regard to Williamina Fleming and her passing, Pickering included a memorial mention in his next \emph{Annual Report of the Director of the Astronomical Observatory of Harvard College for the Year Ending September 30, 1911} \citep{Pickering_1912}, and praised her prodigious talent: 
    
    \begin{quote}
    Mrs. Fleming's record as a discoverer of new stars, stars of the fifth type, and of other objects having peculiar spectra, was unequaled.
    \end{quote}

\section{The Spectrum of H\MakeLowercase{e}$^+$ as a Proving Ground for Bohr's Model of the Atom}
\label{HePlus}

The series of spectral lines associated with Pickering's name played a unique role in confirming the validity of the model of the atom devised in 1913 by Niels Bohr (1885-1962). As noted in Section \ref{Prelude}, Planck's constant made its fourth coming in the Bohr model that would, with the 1916 amendments by Arnold Sommerfeld (1868-1951) and Peter Debye (1884-1966), embody quantum theory until the discovery of quantum mechanics in 1925-1926.   

In the first sequel of his 1913 trilogy on the atomic model, Bohr noted \citep{Bohr_1913He}: 
\begin{quote}
We shall ... see that, [with the] help of the above theory [Bohr's model of the hydrogen atom], we can account naturally for these series of lines [i.e.] the series first observed by Pickering in the spectrum of the star $\zeta$-Puppis, and the set of series recently found by Fowler by experiments with vacuum tubes containing a mixture of hydrogen and helium ... if we ascribe them to helium [with one electron missing].
\end{quote} 

According to Bohr's early version of his atomic model, the wavelength, $\lambda$, of an emission line from state $n_2$ to state $n_1$ (where $n_2>n_1$) of a hydrogenic (one-electron) atom was given by (in modern notation but in the cgs units used by Bohr)
\begin{equation}
\lambda^{-1}=R_Z\left(\frac{1}{n_1^2}-\frac{1}{n_2^2} \right)
\end{equation}
where $R_Z$ is the Rydberg constant pertaining to a one-electron atom with atomic number $Z$,
\begin{equation}
\label{R_Z}
R_Z=\frac{2\pi^2m_eZ^2e^2}{h^3c}
\end{equation}
and $m_e$ is the electron mass, $e$ electron charge, $h$ Planck's constant, and $c$ the speed of light. Upon reading Bohr's paper \citep{Bohr_1913He}, Alfred Fowler (1868-1940) pointed out to Bohr \citep{Fowler_1913a}  a discrepancy between Bohr's treatment of the Pickering series and the positions of the spectral lines of the series as measured in his own laboratory \citep{Fowler_1913} and confirmed by Evan Jenkins Evans at Rutherford's laboratory \citep{Evans_1913}. Indeed, the ratio of the Rydberg constants given by Bohr's formula, Eq. (\ref{R_Z}), for He$^+$ and for atomic hydrogen would be
\begin{equation}
\frac{R_{He^+}}{R_H}=\frac{Z_{He^+}}{Z_H}=4
\end{equation}
whereas for the Pickering series and the Balmer series \citep{Ames_1890}, this ratio came out experimentally as
\begin{equation}
\frac{R_{He^+}}{R_H}\approx4.0016
\end{equation}
Whereupon Bohr went back to the drawing board and replaced in his expression (\ref{R_Z}) for the Rydberg constant the electron mass with the reduced mass, $m_e\mapsto \mu_Z\equiv m_e M_Z/(m_e+M_Z)$, of the nucleus-electron system, as he should have done to begin with \citep{BohrNature_1913}:
\begin{equation}
R_Z=\frac{2\pi^2\mu_Z Z^2e^2}{h^3c}
\end{equation}
With this amendment, the ratio of the Rydberg constants for He$^+$ and H became
\begin{equation}
\label{eq:R_ratio}
\frac{R_{He^+}}{R_H}=\frac{Z_{He^+}\mu_{He^+}}{Z_H \mu_H}=4.00163
\end{equation}
in excellent agreement with the accurate laboratory data, as subsequently acknowledged by Fowler \citep{Fowler_1913}. 

Had the wavelengths of the Pickering series published by Pickering in 1897 been used to determine $R_{He^+}$, the ratio $\frac{R_{He^+}}{R_H}$ would have come out as 4.00101 using the data from the left $\zeta$ column of the table in Ref. \citep{Pickering_1897}\footnote{This is for a value of $R_{He^+}$ 
determined as an average of the values $[R_{He^+}]^{-1}=\lambda\left(n_1^{-2}-n_2^{-2}\right)$ for $\lambda(n_1=4;n_2=11)=4199.2$ \AA, $\lambda(n_1=4;n_2=13)=4027.1$ \AA, $\lambda(n_1=4;n_2=15)=3924.6$ \AA, $\lambda(n_1=4;n_2=17)=3858.7$ \AA, and $\lambda(n_1=4;n_2=19)=3814.7$ \AA ~listed in the left $\zeta$ column of the table in Ref. \citep{Pickering_1897}.} and 4.00011 using the data from the right $\zeta$ column of the table in Ref. \citep{Pickering_1897}.\footnote{This is for a value of $R_{He^+}$ 
determined as an average of the values $[R_{He^+}]^{-1}=\lambda\left(n_1^{-2}-n_2^{-2}\right)$ for $\lambda(n_1=4;n_2=11)=4201.6$ \AA, $\lambda(n_1=4;n_2=13)=4026.5$ \AA, $\lambda(n_1=4;n_2=15)=3924.9$ \AA, $\lambda(n_1=4;n_2=17)=3858.6$ \AA, and $\lambda(n_1=4;n_2=19)=3817.2$ \AA ~listed in the right $\zeta$ column of the table in Ref. \citep{Pickering_1897}.}
Interestingly enough, the blue $\lambda=4686$ \AA ~ line corresponding to the transition between the $n_1=3$ and $n_2=4$ levels was not reported in communications from the HCO in the 1890s. This line typifies the Pickering series as observed in laboratory discharge-tube spectra \citep{Fowler_1912}.

In 1915, Bohr provided further comments on the differences between the Balmer and Pickering series and strengthened his case by referring to 1914 experimental data on the ionization potentials of hydrogen and helium \citep{Bohr_1915}. 
 
The explanation of the Pickering series was a major triumph for Bohr's theory. Thereby, Bohr's theory proved capable of not just reproducing the well-known and well-established, such as the Balmer series or the Rydberg constant for hydrogen, but it also demonstrated its predictive and explanatory power. In particular, the accurate rendition of the ratio, Eq. (\ref{eq:R_ratio}), of the Rydberg constants for He$^+$ and hydrogen had made a deep impression. When Albert Einstein (1879-1955) heard about it at a 1913 conference, he exclaimed \citep{Stachel_2001}, pp. 369-370:
\begin{quote}
Then the frequency of the light does not depend at all on the frequency of the electron ... this is an enormous achievement. The theory of Bohr must then be right. 
\end{quote}
In his 1951 reflections on the advent of quantum theory, Einstein took the opportunity to pay tribute to Bohr's achievement once more \citep{Einstein_1951}:
\begin{quote}
[W]ithout having a substitute for classical mechanics, I could nevertheless see to what kind of consequences
this law of temperature-radiation [black body radiation law] leads for the photoelectric effect and for other related phenomena of the transformation of radiation-energy, as well as for the specific heat of (especially) solid bodies. All my attempts, however, to adapt the theoretical foundation of physics to this (new type of) knowledge failed completely. It was as if the ground had been pulled out from under one, with no firm foundation to be seen anywhere, upon which one could have built. That this insecure and contradictory foundation was sufficient to enable a man of Bohr's unique instinct and tact to discover the major laws of the spectral lines and of the electron-shells of the atoms, together with their significance for chemistry, appeared to me like a miracle -- and appears to me as a miracle even today. This is the highest form of musicality in the sphere of thought.
\end{quote}

Bohr's model of the hydrogenic (i.e., one-electron) atom was amended in 1916 by Arnold Sommerfeld (1868-1951) who made use of relativistic Old Quantum Theory to explain the fine structure of the H and He$^+$ spectra and to provide the means to meet the challenge of unriddling the anomalous Zeeman effect \citep{som1916}, \citep{Paschen_1916}, and \citep{HSB_Frie_2023}. Sommerfeld and, independently, Peter Debye (1884-1966) \citep{deb1916}, concluded that not just the magnitudes of the electronic orbital angular momenta but also the spatial orientations of the electronic orbits with respect to an external magnetic field are quantized. The Bohr-Sommerfeld-Debye model was subjected in 1922 to a non-spectroscopic test, the Stern-Gerlach experiment \citep{gs1922b}, which confirmed the existence of space quantization and thus ruled unequivocally in favor of quantum mechanics as epitomized by the Bohr-Sommerfeld-Debye model \citep{Friedrich_2023}. 

Bohr's model of the atom thus continued playing the role of a touchstone on the path to quantum theory until its advent in 1925-1927. We add that a definitive treatment of the fine structure of hydrogenic atoms as due to spin-orbit coupling and relativistic effects was provided by Sommerfeld in 1940 \citep{Sommerfeld_1940} and reviewed by numerous authors, including Condon and Shortley \citep{Condon_Shortley_1951}.

	\section{Epilogue}
	\label{Conclusions}
	
	Cecilia Payne-Gaposchkin, in her 1925 PhD thesis written at the Harvard College Observatory, provided this assessment of the significance of the test of Bohr's model afforded by the measurement of the Pickering series \citep{Cecilia_1925}, p. 14:
\begin{quote}
The detection and resolution of the alternate components of [the Pickering] series, which fall very near to the Balmer lines of hydrogen in the spectra of the hottest stars, and the consequent derivation of the Rydberg constant for helium [plus] ... represents an astrophysical contribution to pure physics which is of the highest importance.
\end{quote}

However, the spectral properties of He$^+$ proved to be of great consequence also for astronomy itself, following the discovery of the enabling role of He$^+$ in the pulsation of the Cepheids and of other variable stars populating chiefly the Instability Strip of the Herzsprung-Russell Diagram \citep{Eddington_1917}, \citep{Vitense_1951}, \citep{Zhevakin_1953}, and \citep{Cox_1963}: In the dimmest (and hottest) phase of the Cepheid cycle, the radiation emanating from the stellar interior is scattered and absorbed by He$^+$ in the stellar envelope, leading to the envelope's expansion and, thus, cooling. At the coolest point of the cycle, the passage of the thermal radiation from the stellar interior through the envelope is the least impeded and so the star shines the brightest. The workings of variable stars could thus be likened to a Carnot cycle with He$^+$  playing the role of the intake/exhaust valve.\footnote{Henrietta Leavitt (1868-1921) discovered in 1908 at the HCO the nearly linear relationship between the maximum luminosity and the period of the Cepheids \citep{Leavitt_1908}, which would become instrumental for establishing Hubble's Law from observational data in 1929 \citep{Hubble_1929}.}

Moreover, in the 1990s, the chemical properties of the He$^+$ ion would be recognized as key to astrochemistry, through the ion's role in the kinetically rather than thermodynamically controlled chemistry of the interstellar medium. In the 1970s, laboratory experiments conducted in conjunction with quantum-chemical calculations \citep{Mahan_1975} showed that while He$^+$ does not react with the most abundant interstellar molecule, H$_2$, it does react more than willingly with the second most abundant molecule, CO, 
\begin{equation}
\mathrm{He^+ + CO \rightarrow He + C^+ + O}
\end{equation} 
yielding C$^+$ in essentially every He$^+$ + CO collision \citep{Klemperer_1995} and \citep{Klemperer_2006}. This enhances the concentration of the C$^+$ ion by the He/CO abundance ratio, i.e., by about a factor of a thousand. The abundant C$^+$ ion reacts only reluctantly with the prevalent H$_2$ molecules and so is spared for its avid reactions with methane and acetylene, giving rise to a reaction sequence responsible for the formation of many organic compounds. As Dudley Herschbach put it \citep{Herschbach_1999}:
\begin{quote}
The paradoxical irony is that the mutual distaste of the simplest inorganic species, He$^+$ and H$_2$, [for one another] gives rise to the proliferation of complex organic molecules in the cold interstellar clouds.
\end{quote}

Edward Pickering's approach of collecting hosts of quality data in the face of ignorance resulted in many unexpected payoffs. Williamina Fleming's sifting through the data uncovered precious nuggets that keep dazzling us to this day.
\newpage
			
	\section*{Acknowledgements}
		For assistance with images, we wish to thank Thomas Burns, Curator of Astronomical Photographs, Harvard College Observatory along with the Staff of the Harvard College Observatory Plate Stacks. We also express our appreciation for help with searching for plates to Lisa Bravata, Curatorial Assistant, Harvard College Observatory Plate Stacks.
We would also like to express our thanks to Sara J. Schechner, the David P. Wheatland Curator of the Collection of Historical Instruments, Harvard University, and to Sara Frankel, Collections Manager, Collection of Historical Instruments, Harvard University. 
We also thank the Staff of the John G. Wolbach Library at the Center for Astrophysics | Harvard \& Smithsonian for their support. 
B.F. thanks Georgene \& Dudley Herschbach for discussions related to this Chapter, and John Doyle and Hossein Sadeghpour for their hospitality during his stay in 2021-2023 at Harvard Physics and at the Institute for Theoretical Atomic, Molecular, and Optical Physics (ITAMP) of the Center for Astrophysics $\vert$ Harvard \& Smithsonian.


\begin{thebibliography}{99}

\bibitem[Ames(1890)]{Ames_1890} Ames, J.S. (1890). `V. On some gaseous spectra: hydrogen, nitrogen.' \emph{The London, Edinburgh, and Dublin Philosophical Magazine and Journal of Science}, 30, 48--58.
\bibitem[Anderson and Roughley(2018)]{ScotlandMortality} Anderson, M. and Roughley, C. (2018) `Causes of Death' in \emph{Scotland's Populations from the 1850s to Today}. Oxford: Oxford University Press.
\bibitem[Bailey(1895)]{Bailey_1895} Bailey, S.I. (1895). `Photometric Observations of Southern Stars. Chapter 1: History of the Expedition.' \emph{Annals of the Astronomical Observatory of Harvard College}, 34, 22.
\bibitem[Bailey(1931)]{Bailey_1931} Bailey, S.I. (1931). \emph{The History and Work of Harvard Observatory, 1839-1927}. New York and London: McGraw-Hill Book Company.
\bibitem[Bailey(1932)]{Bailey_1932} Bailey, S.I. (1932). `Biographical Memoir of Edward Charles Pickering 1846-1919.' \emph {National Academy of Sciences Biographical Memoirs}, 15, 169--189.
\bibitem[Balmer(1885)]{Balmer_1885} Balmer, J. (1885). `Notiz \"uber die Spektrallinien des Wasserstoffs.' \emph{Verhandlungen der Naturforscher Gesellschaft zu Basel}, 7, p. 548.
\bibitem[Barker(1888)]{Barker_1888} Barker, G.F. (1888). `Biographical Memoir of Henry Draper 1837-1882.' \emph{National Academy of Sciences Biographical Memoirs}, 81--139.
\bibitem[BIAU(2022)]{WilliaminaAsteroid} BIAU (2022). \emph{Bulletin of the International Astronomical Union}, 2, 5.
\bibitem[Bohr(1913a)]{Bohr_1913He} Bohr, N. (1913a). `I. on the constitution of atoms and molecules.' \emph{The London, Edinburgh, and Dublin Philosophical Magazine and Journal of Science}, 26, 1--25.
\bibitem[Bohr(1913b)]{BohrNature_1913} Bohr, N. (1913b). `The Spectra of Helium and Hydrogen.'  \emph{Nature}, 92, 231--232.
\bibitem[Bohr(1915)]{Bohr_1915} Bohr, N. (1915). `The spectra of hydrogen and helium.'  \emph{Nature}, 95, 6--7.
\bibitem[Boston Herald(1899)]{Herald_1899_1} Boston Herald (1899). `Harvard Honors Women.' \emph{Boston Herald}, January 16.
\bibitem[Boston Herald(1911)]{Herald_1911} Boston Herald (1911). `Leading Woman in Science Dies in a Hospital: Mrs. Fleming, Harvard's Famous Astronomer, 30 Years at Observatory; Honored by Royal Society; Discovered More Stars Than All Others in Profession in 200 Years; Original in Her Methods; First to Detect Approach of Halley's Comet Only Two Years Ago.' \emph{ Boston Herald}, May 22.
\bibitem[Cannon(1911)]{Cannon_1911} Cannon, A.J. (1911). `Williamina Paton Fleming.' \emph{Astrophysical Journal}, 34, 315--316.
\bibitem[Cannon(1912)]{AJC_1912Class} Cannon, A.J. (1912). `Classification of 1,477 Stars by Means of their Photographic Spectra.' \emph{Annals of the Astronomical Observatory of Harvard College}, 56, 66.
\bibitem[Cannon(1915)]{AJC_1915} Cannon, A.J. (1915). `The Henry Draper Memorial.' \emph{Journal of the Royal Astronomical Society of Canada}, 9, 205--206.
\bibitem[Cannon(1919)]{AJC_1919} Cannon, A.J. (1919). `Edward Charles Pickering.' \emph{Popular Astronomy}, 27, 177--182.
\bibitem[Cannon(1994)]{Draper} Cannon, A.J. (1994). `Henry Draper Catalogue and Extension.' 1994.
\bibitem[Clerke(1896)]{Clerke_1896} Clerke, M.A. (1896). `New Views about Mars.' \emph{Edinburgh Review}, 184, 184.
\bibitem[Condon and Shortley(1951)]{Condon_Shortley_1951} Condon, E. and Shortley, G. (1951). \emph{The Theory of Atomic Spectra}. Cambridge: Cambridge University Press.
\bibitem[Cox(1963)]{Cox_1963} Cox, J.P. (1963). `On Second Helium Ionization as a Cause of Pulsational Instability in Stars.' \emph{Astrophysical Journal}, 138, 487.
\bibitem[Debye(1916)]{deb1916} Debye, P. (1916). `Quantenhypothese und Zeeman-Effekt.' \emph{Physikalische Zeitschrift}, 17, 507--512.
\bibitem[Donaghe(1898)]{Donaghe_1898} Donaghe, H.R. (1898). `Photographic Flashes from Harvard Observatory.' \emph{Popular Astronomy}, 6, 483.
\bibitem[DPedia(2023)]{PickeringMartianCrater} DPedia (2023). \emph{Pickering (Martian Crater)}. DBpedia.
\bibitem[Eddington(1917)]{Eddington_1917} Eddington, A.S. (1917). `The pulsation theory of Cepheid variables.' \emph{The Observatory}, 40, 290--293.
\bibitem[Einstein(1951)]{Einstein_1951} Einstein, A. (1951). `The advent of the quantum theory,' \emph{Science}, 113, 82--84.
\bibitem[Evans(1913)]{Evans_1913} Evans, E.J. (1913). `The Spectra of Helium and Hydrogen.' \emph{Nature}, 92, 5.
\bibitem[Fiss(2023)]{Fiss_2023} Fiss, A. (2023). `For Computing Is Our Duty,' in \emph{Algorithmic Modernity} (M.~G. Ames and M.~Mazzotti, eds.). Oxford: Oxford University Press.
\bibitem[Fleming(1893a)]{Fleming_1893c} Fleming, W. (1893a) `A Field for Woman's Work in Astronomy.' \emph{Astronomy and Astrophysics}, 12, 687.
\bibitem[Fleming(1893b)]{Fleming_1893b} Fleming, W. (1893b). `A Field for Woman's Work in Astronomy.' \emph{Astronomy and Astrophysics}, 12, 685--686.
\bibitem[Fleming(1893c)]{Fleming_1893a} Fleming, W. (1893c). `A Field for Woman's Work in Astronomy.' \emph{Astronomy and Astrophysics}, 12, 683--689.
\bibitem[Fleming(1896a)]{Fleming_1896_Logbook} Fleming, W. (1896a). `Variable Stars.' \emph{phaedra0809}, Cambridge, MA: John G. Wolbach Library, Harvard College Observatory, Project PHAEDRA. Harvard College Observatory observations, logs, instrument readings, and calculations, 9, 146--147, 1896/1897.
\bibitem[Fleming(1900)]{Fleming_1900} Fleming, W.P.S. (1900). \emph{Journal of Williamina Paton Fleming, 1900 Mar. 1 -- Apr. 18; Curator of Astronomical Photographs, Harvard College Observatory}. Harvard College Observatory. Chest of 1900.
\bibitem[Fleming(1907)]{Fleming_1907Variable} Fleming, W.P. (1907). `Photographic Study of Variable Stars.' \emph{Annals of the Astronomical Observatory of Harvard College}, 47, 2.
\bibitem[Fleming(1912)]{Fleming_1912} Fleming, W.P. (1912). `Stars Having Peculiar Spectra.' \emph{Annals of the Astronomical Observatory of Harvard College}, 56, 178.
\bibitem[Fowler(1912)]{Fowler_1912} Fowler, A. (1912). `Observations of the principal and other series of lines in the spectrum of Hydrogen (Plates 2-4).'  \emph{Monthly Notices of the Royal Astronomical Society}, 73, 62--72.
\bibitem[Fowler(1913a)]{Fowler_1913a} Fowler, A. (1913a). `The spectra of helium and hydrogen.' \emph{Nature}, 92(2291), 95--96.
\bibitem[Fowler(1913b)]{Fowler_1913} Fowler, A. (1913b). `Letter to the Editor.' \emph{Nature}, 92, 232--233.
\bibitem[Friedrich(2023)]{Friedrich_2023} Friedrich, B. (2023). `A Century Ago the Stern-Gerlach Experiment Ruled Unequivocally in Favor of Quantum Mechanics.' \emph{Israel Journal of Chemistry}, 63, e2023000.
\bibitem[Gerlach and Stern(1922)]{gs1922b} Gerlach, W. and Stern, O. (1922). `Der experimentelle Nachweis der Richtungsquantelung im Magnetfeld.' \emph{Zeitschrift f\"ur Physik}, 9, 349--352.
\bibitem[Grier(2013)]{Grier_2013} Grier, D.A. (2013). \emph{When Computers Were Human}. Princeton: Princeton University Press.
\bibitem[Haas(1910)]{Haas_1910} Haas, A.E. (1910). `\"Uber eine neue theoretische Bestimmung des elektrischen Elementarquantums und des Halbmessers des Wasserstoffatoms.' \emph{Physikalische Zeitschrift}, 11, 537.
\bibitem[Hearnshaw(2014)]{Hearnshaw_2014} Hearnshaw, J. (2014).  \emph{Analysis of Star Light: Two Centuries of Astronomical Spectroscopy, 2nd Edition}. Cambridge: Cambridge University Press.
\bibitem[Herschbach(1999)]{Herschbach_1999} Herschbach, D.R. (1999). `Chemical Physics: Molecular Clouds, Clusters, and Corrals' in \emph{More Things in Heaven and Earth. A Celebration of Physics at the Millennium, Volume II} (B.~Bederson, ed.), pp.~693--705. Heidelberg: Springer.
\bibitem[Hoffleit(1991)]{Hoffleit_1991} Hoffleit, D. (1991). `Evolution of the Draper Memorial.' \emph{Vistas in Astronomy}, 34, p. 118.
\bibitem[Hoffleit(1997)]{Hoffleit_1997} Hoffleit, D. (1997). `History of the Discovery of Mira Stars.' \emph{Journal of the American Association of Variable Star Observers}, 25, 132.
\bibitem[Holberg(2007)]{Holberg_2007} Holberg, J. (2007). \emph{Sirius: the Brightest Diamond in the Night Sky}. Berlin, New York, Chichester: Springer.
\bibitem[Hubble(1929)]{Hubble_1929} Hubble, E. (1929). `A Relation between Distance and Radial Velocity among Extra-Galactic Nebulae.' \emph{Proceedings of the National Academy of Sciences}, 15, 168--173.
\bibitem[Hughes(2012)]{Hughes_2012} Hughes, S. (2012). \emph{Catchers of the Light. A History of Astrophotography}. ArtDeCiel Publishing.
\bibitem[IAU(2010)]{FlemingLunarCrater} IAU (2010). \emph{Gazetteer of Planetary Nomenclature}. International Astronomical Union (IAU).
\bibitem[IAU(2010)]{PickeringLunarCrater} IAU (2010). \emph{Gazetteer of Planetary Nomenclature}. International Astronomical Union (IAU).
\bibitem[Johnson(2005)]{Johnson_2005} Johnson, G. (2005). \emph{Miss Leavitt's Stars. The Untold Story of the Woman Who Discovered How to Measure the Universe}. New York: Norton.
\bibitem[Jones and Boyd(1971)]{Zaban_1971} Jones, B.Z. and Boyd, L.G. (1971). \emph{Harvard College Observatory. The First Four Directorships}. Cambridge, MA: Harvard University Press.
\bibitem[Klemperer(1995)]{Klemperer_1995} Klemperer, W. (1995). `Some Spectroscopic Reminiscences.' \emph{Annual Review of Physical Chemistry}, 46(1), 1--28. PMID: 22559036.
\bibitem[Klemperer(2006)]{Klemperer_2006} Klemperer, W.A. (2006). `Interstellar Chemistry.' \emph{Proceedings of the National Academy of Sciences}, 103(33), 12232--12234.
\bibitem[Kragh(2002)]{Kragh_2002} Kragh, H. (2002). \emph{Quantum Generations: A History of Physics in the Twentieth Century}. Heidelberg: Princeton University Press, 2002.
\bibitem[LaFortune(2001)]{LaFortune_2001} LaFortune, K. (2001). \emph{Women at the Harvard College Observatory, 1877-1919:``Women's Work,'' the ``New'' Sociality of Astronomy and Scientific Labor. M.A. Thesis}. Indiana: University of Notre Dame.
\bibitem[Lankford(1997)]{Lankford_1997} Lankford, J. (1997). \emph{American Astronomy: Community, Careers, and Power}. Chicago: University of Chicago Press.
\bibitem[Leavitt(1908)]{Leavitt_1908} Leavitt, H.S. (1908). `1777 variables in the Magellanic Clouds.' \emph{Annals of Harvard College Observatory}, 60, 87--108.
\bibitem[Lewis(1885)]{Lewis_1885} Lewis, T. (1885). `Harvard Photometry.' \emph{Observatory}, 8, 49.
\bibitem[Mahan(1975)]{Mahan_1975} Mahan, B.H. (1975). `Electronic structure and chemical dynamics.' \emph{Accounts of Chemical Research}, 8, 55--61.
\bibitem[Massachusetts Naturalization Records(1900)]{Naturalization} Massachusetts Naturalization Records (1900). \emph {Massachusetts; Naturalization Records - Originals, 1906-1929 for Williamina Fleming}.
\bibitem[Nature(1912)]{Nature_1912} Nature (1912). `New Hydrogen Spectra.' \emph{Nature}, 90, 466--467.
\bibitem[New York Times(1888)]{NYT_1888} New York Times (1888). `Nebula in Orion; Prof. Henry Draper's Photographs of the Spectrum.' \emph{The New York Times}, p.~9, 2 April 1888.
\bibitem[Nicholson(1911)]{Nicholson_1911} Nicholson, M. (1911). `A Structural Theory of the Chemical Elements.' \emph{Philosophical Magazine}, 22, 864.
\bibitem[Pais(1986)]{Pais_1986} Pais, A. (1986). \emph{Inward Bound}. Oxford: Oxford University Press.
\bibitem[Paschen(1916)]{Paschen_1916} Paschen, F. (1916). `Bohr's Heliumlinien.'  \emph{Annalen der Physik}, 355(16), 901--940.
\bibitem[Payne-Gaposchkin(1979)]{CPG_1979} Payne-Gaposchkin, C. (1979). \emph{Dyer's Hand: An Autobiography}. Privately printed, 1979.
\bibitem[Payne(1925)]{Cecilia_1925} Payne, C.H. (1925).  \emph{Stellar Atmospheres; a Contribution to the Observational Study of High Temperature in the Reversing Layers of Stars}. PhD Thesis, The Observatory, Cambridge, Massachusetts.
\bibitem[Pickering and Fleming(1890)]{Pick_Flem_1890} Pickering, E.C. and Fleming, M. (1890). `Stars Having Peculiar Spectra.' \emph{Sidereal Messenger}, 379, 379--380.
\bibitem[Pickering and Fleming(1891)]{Pick_Flem_1891} Pickering, E.C. and Fleming, M. (1891). `Stars Having Peculiar Spectra. New Variable Stars in Aquarius, Delphinus and Camelopardalus.' \emph{Astronomische Nachrichten}, 127, 5.
\bibitem[Pickering(1869)]{Pickering_1869} Pickering, E.C. (1869). \emph{Plan of the Physical Laboratory of the Massachusetts Institute of Technology}. Boston: Press of A.A. Kingman, Museum of the Boston Society of Natural History.
\bibitem[Pickering(1881a)]{Pickering_1881} Pickering, E.C. (1881a). `New Variable Star in Puppis.' \emph{Astronomische Nachrichten}, 100, 13--14.
\bibitem[Pickering(1881b)]{Pickering_Schreiben_1881} Pickering, E.C. (1881b). `Schreiben des Herrn Professor Edw. C. Pickering an den Herausgeber.' \emph{Astronomische Nachrichten}, 99, 375--378.
\bibitem[Pickering(1882)]{Pickering_1882} Pickering,  E.C. (1882). \emph{Plan for Securing Observations of the Variable Stars}. Harvard College Observatory Papers, Vol. 1, 1882.
\bibitem[Pickering(1882Stars)]{Pickering_1882Stars} Pickering, E.C. (1882). `Stars with Peculiar Spectra Discovered at the Astronomical Observatory of Harvard College.' \emph{Astronomische Nachrichten}, 101, 73--74.
\bibitem[Pickering(1896a)]{Pickering_1896Crux} Pickering, E.C. (1896a). `Stars Having Peculiar Spectra. New Variable Stars in Crux and Cygnus.'  \emph{Harvard College Observatory Circular No. 12}, 1--2.
\bibitem[Pickering(1896b)]{Pickering_1896Nachr} Pickering, E.C. (1896b). `Stars Having Peculiar Spectra. New Variable Stars in Crux and Cygnus.'  \emph{Astronomische Nachrichten}, 142, 87--90.
\bibitem[Pickering(1896c)]{Pickering_1896Astr} Pickering, E.C. (1896c). `Stars Having Peculiar Spectra. New Variable Stars in Crux and Cygnus.'  \emph{Astrophysical Journal}, 4, 369--370.
\bibitem[Pickering(1896d)]{Pickering_1896} Pickering, E.C. (1896d). `Stars having peculiar spectra. New variable Stars in Crux and Cygnus.'  \emph{Astronomische Nachrichten}, 142(6), 87--90.
\bibitem[Pickering(1897a)]{Pickering_1897Southern} Pickering, E.C. (1897Southern). `Southern Double Stars.' \emph{Harvard College Observatory Circular No. 18}, 1.
\bibitem[Pickering(1897b)]{Pickering_1897Circ} Pickering, E.C. (1897b). `Spectrum of $\zeta$ Puppis.'  \emph{Harvard College Observatory Circular No. 16}, 1--2.
\bibitem[Pickering(1897c)]{Pickering_1897Nachr} Pickering, E.C. (1897c). `Spectrum of $\zeta$ Puppis.'  \emph{Astronomische Nachrichten}, 142, 399--402.
\bibitem[Pickering(1897d)]{Pickering_1897Astr} Pickering, E.C. (1897d). `Spectrum of $\zeta$ Puppis.'  \emph{Astrophysical Journal}, 5, 92--94.
\bibitem[Pickering(1897e)]{Pickering_1897} Pickering, E.C. (1897e). `The spectrum of zeta Puppis.' \emph{Astrophysical Journal}, 5(5), 92--94.
\bibitem[Pickering(1901)]{Pickering_1901Circ} Pickering, E.C. (1901). `Spectrum of $\zeta$ Puppis.' \emph{Harvard College Observatory Circular No. 55}, 1--3.
\bibitem[Pickering(1912)]{Pickering_1912} Pickering, E.C. (1912). \emph{Sixty-Sixth Annual Report of the Director of The Astronomical Observatory of Harvard College for the Year Ending September 30, 1911}. Cambridge, Mass.: The University, 1912.
\bibitem[Pickering(1915)]{Pickering_1915} Pickering, E.C. (1915). `Objective Prism.' \emph{Popular Astronomy}, 23, 489.
\bibitem[Plotkin(1990)]{Plotkin_1990} Plotkin, H. (1990). `Edward Charles Pickering.' \emph{Journal for the History of Astronomy}, xxi, 48.
\bibitem[Robotti(1983)]{Robotti_1983} Robotti, N. (1983). `The Spectrum of $\zeta$ Puppis and the Historical Evolution of Empirical Data.'  \emph{Historical Studies in the Physical Sciences}, 14(1), 123--145.
\bibitem[Russell(1914)]{Russell_1914} Russell, H.N. (1914). `Relations Between the Spectra and other Characteristics of the Stars,'  \emph{Popular Astronomy}, 22, 276.
\bibitem[Rydberg(1890)]{Rydberg_1890} Rydberg, J.R. (1890). `On the structure of the line-spectra of the chemical elements.'  \emph{The London, Edinburgh, and Dublin Philosophical Magazine and Journal of Science}, 29(179), 331--337.
\bibitem[Schilbach(2008)]{Schilbach_2008} Schilbach, E. and R\"oser, S. (2008). `On the origin of field O-type stars.'  \emph{Astronomy \& Astrophysics}, 489(1), 105--114.
\bibitem[Schmadel(2007)]{Schmadel_2007} Schmadel, L.D. (2007). `(784) Pickeringia' in \emph{Dictionary of Minor Planet Names}, pp.~74--74, Berlin, Heidelberg: Springer.
\bibitem[Schmidt-B\"ocking, Gruber, and Friedrich(2022)]{HSB_Frie_2023} Schmidt-B\"ocking, H. and Gruber, G. and Friedrich, B. (2022). `One hundred years ago Alfred Land\'e unriddled the Anomalous Zeeman Effect and presaged electron spin.' \emph{Physica Scripta}, 98, 014005.
\bibitem[Sheehan(2015)]{Sheehan_2015} Sheehan, W. (2015).  \emph{Galactic Encounters: Our Majestic and Evolving Star System, from the Big Bang to Time's End}. New York: Springer.
\bibitem[Shindell(2023)]{Shindell_2023} Shindell, M. (2023). \emph{For the Love of Mars}. Chicago: Chicago University Press.
\bibitem[Sobel(2016)]{Sobel_2016} Sobel, D. (2016). \emph{The Glass Universe. How the Ladies of the Harvard Observatory Took the Measure of the Stars}. New York: Viking.
\bibitem[Sommerfeld(1916)]{som1916} Sommerfeld, A. (1916). `Zur Quantentheorie der Spektrallinien,' \emph{Annalen der Physik}, 51, 125--167.
\bibitem[Sommerfeld(1940)]{Sommerfeld_1940} Sommerfeld, A. (1940). `Zur Feinstruktur der Wasserstofflinien. Geschichte und gegenw{\"a}rtiger Stand der Theorie.' \emph{Naturwissenschaften}, 28, 417--423.
\bibitem[Stachel(2001)]{Stachel_2001} Stachel, J. (2001).  \emph {Einstein from `B' to `Z'}. Boston: Birkh\"auser.
\bibitem[Turner(1912)]{Turner_1912} Turner, H.H. (1912). `Report of the Council to the Ninety-Second Annual General Meeting: Mrs. Fleming.' \emph{Monthly Notices of the Royal Astronomical Society}, 72, 261--264.
\bibitem[Upton(2001)]{100CPG_2001} Upton, W. (2001). `HCO Pinafore' in \emph{The Starry Universe. The Cecilia Payne-Gaposchkin Centenary} (Davis Philip, A. G. and Koopmann, Rebecca A., eds.), Schenectady, New York, U.S.A.: L. Davis Press, 2001.
\bibitem[Vitense(1951)]{Vitense_1951} Vitense, E. (1951). `Der Aufbau der Sternatmosph\"aren IV. Teil, Kontinuierliche Absorption und Streuung als Funktion von Druck und Temperatur.' \emph{Zeitschrift f\"ur Astrophysik}, 28, 81--112.
\bibitem[Welther(1982)]{Welther_1982} Welther, B.L. (1982). `Pickering's Harem.' \emph{Isis}, 73, 94--94.
\bibitem[Wiescher(2021)]{Wiescher_2021} Wiescher, M. (2021). \emph{Arthur E. Haas--The Hidden Pioneer of Quantum Mechanics. A Biography}. Cham: Springer Nature.
\bibitem[Zhevakin(1953)]{Zhevakin_1953} Zhevakin, S. (1953). `On the Theory of the Cepheids.' \emph{Russian Astronomical Journal}, 30, 161.

\end{thebibliography}
\end{document}